 \documentclass[final,5p,times,twocolumn]{elsarticle}

\usepackage{amssymb}

\usepackage[utf8]{inputenc}
\usepackage{color}
\usepackage[english]{babel}
\usepackage{url}
\usepackage{amsmath}
\usepackage{amssymb}
\usepackage{graphicx}
\usepackage{esint}
\usepackage[unicode=true]{hyperref}
\usepackage{adjustbox}
\usepackage{citesort}
\usepackage{subcaption}

\newcounter{bla}

\usepackage{amsmath, amssymb, graphics, setspace}
\newcommand{\mathsym}[1]{{}}
\newcommand{\unicode}[1]{{}}

\newcommand\cmd[1]{{\tt #1}}

\allowdisplaybreaks

\newenvironment{myquotation}{\setlength{\leftmargini}{-0.5em}\color{blue}\quotation}{\endquotation}
\def\lg{\textless \textgreater}
\newcommand{\ts}{\rule{.4em}{.4pt}}  
\def\pmb#1{{\color{blue} \small \bf #1}}
\def\text#1{#1}

\def\citep#1{\cite{#1}}
\def\citet#1{\cite{#1}}
\long\def\junk#1{\null}

\journal{Computer Physics Communications}

\begin{document}

\newcommand{\Mane}{\cmd{ManeParse}}
\newcommand{\fortran}{\cmd{FORTRAN}}
\newcommand{\lha}{\cmd{LHAPDF6}}
\newcommand{\cteq}{CTEQ}
\newcommand{\infoFile}{\cmd{.info}}
\newcommand{\dataFile}{\cmd{.dat}}
\newcommand{\cteqFile}{\cmd{.pds}}
\newcommand{\mstw}{MSTW}
\newcommand{\nnpdf}{NNPDF}
\newcommand{\mma}{\cmd{Mathematica}}
\newcommand{\cpp}{\cmd{C$++$}}
\newcommand{\hepforge}{\cmd{HEPForge}}
\newcommand{\perl}{Perl}

\newcommand\crule[1][black]{\textcolor{#1}{\rule{0.5cm}{0.5cm}}} 
\definecolor{math1}{rgb}{0.368417, 0.506779, 0.709798} 
\definecolor{math2}{rgb}{0.880722, 0.611041, 0.142051} 
\definecolor{math3}{rgb}{0.560181, 0.691569, 0.194885} 
\definecolor{math4}{rgb}{0.922526, 0.385626, 0.209179} 
\definecolor{math5}{rgb}{0.528488, 0.470624, 0.701351} 
\definecolor{math6}{rgb}{0.772079, 0.431554, 0.102387} 
\definecolor{math7}{rgb}{0.363898, 0.618501, 0.782349} 
\definecolor{math8}{rgb}{1, 0.75, 0} 
\definecolor{math9}{rgb}{0.647624, 0.37816, 0.614037} 
\definecolor{math10}{rgb}{0.571589, 0.586483, 0.} 

\graphicspath{{./images/}}

\begin{frontmatter}



  \title{\Mane: A \mma\ reader for Parton Distribution Functions  }
  \tnotetext[t1]{Published in: {\bf Computer Physics Communications 216 (2017) 126-137} \\}


\author[a]{D.~B.~Clark\corref{author1}}
\ead{dbclark\mbox{}@\mbox{}smu.edu}
\author[a]{E.~Godat\corref{author2}}
\ead{egodat\mbox{}@\mbox{}smu.edu}
\author[a]{F.~I.~Olness\corref{author3}}
\ead{olness\mbox{}@\mbox{}smu.edu}

\address[a]{Department of Physics, Southern Methodist University, Dallas, TX
75275, USA}

\begin{abstract}
Parton Distribution Functions (PDFs) are essential non-perturbative
inputs for calculation of any observable with hadronic initial states.
These PDFs are released by individual groups as discrete grids
as a function of the Bjorken-$x$ and energy scale $Q$. The 
LHAPDF project maintains a repository of PDFs from various groups in a new
standardized \lha\ format, additionally  older formats such as the
\cteq\ PDS grid format are still in use.
\Mane\ is a package that provides access to PDFs within \mma\
to facilitate calculation and plotting. The program is self-contained
so there are no external links to any \fortran, {\tt C}\ or \cpp\ programs.
The package includes the option to use the built-in \mma\ interpolation
or a custom cubic Lagrange interpolation routine which allows for
flexibility in the extrapolation (particularly at small $x$-values).
\Mane\ is fast enough to enable simple calculations (involving even
one or two integrations) in the \mma\
framework. 

\end{abstract}

\begin{keyword}
QCD; Mathematica; Parton Distribution Functions; PDF; PDFs; Hadron
collider; PDF Errors; Hadronic Cross Section
\end{keyword}

\end{frontmatter}



\newpage
\noindent
{\bf PROGRAM SUMMARY}

\begin{small}
\noindent
{\em Program Title: \Mane}                                          \\
{\em Licensing provisions(please choose one): MIT}                                   \\
{\em Programming language: \mma}                                   \\

\noindent
{\em Nature of problem(approx. 50-250 words):}\\
  PDFs are currently read and interpolated via a \fortran\ or \cpp\
interface. No~method exist to read the \lha\ or \cteq\ PDFs directly
in \mma.\\
\\ \noindent
{\em Solution method(approx. 50-250 words):}\\
  A \mma\ package reads in \lha\ and \cteq\ PDF files. The
PDFs are parsed into a three-dimensional array in Bjorken-$x$, scattering
energy $Q$, and parton flavor, and are stored in memory. Provided
functions give access to the PDF, the PDF uncertainty, the PDF correlations,
and the parton-parton Luminosities. The \lha\ info files are converted
from YAML format into Mathematica rules. \\
   \\

\end{small}

\newpage



\section{What is \Mane? \label{sec:intro}}

\begin{figure}[t] 
\centering{}
\includegraphics[width=0.45\textwidth]{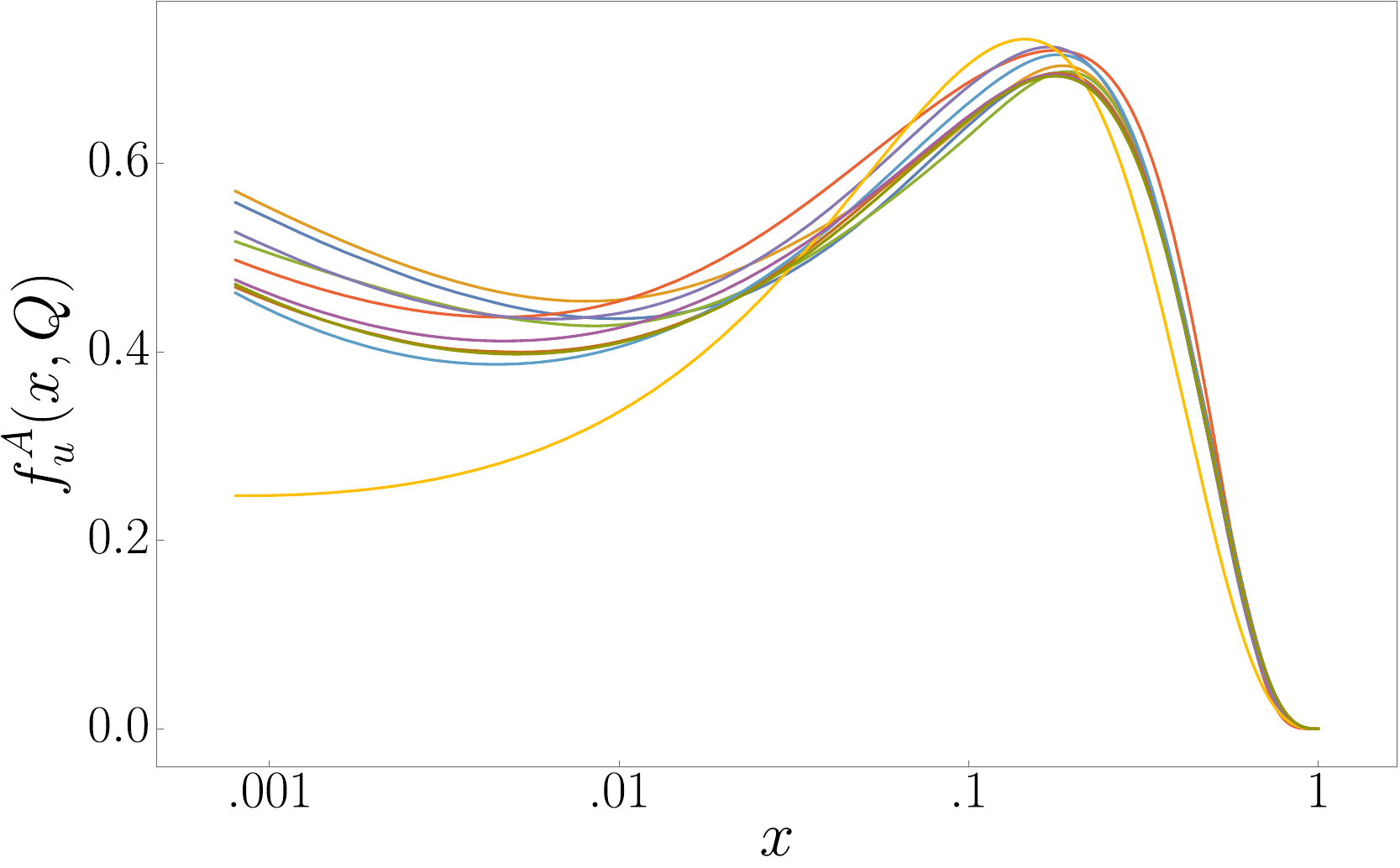}
\includegraphics[width=0.45\textwidth]{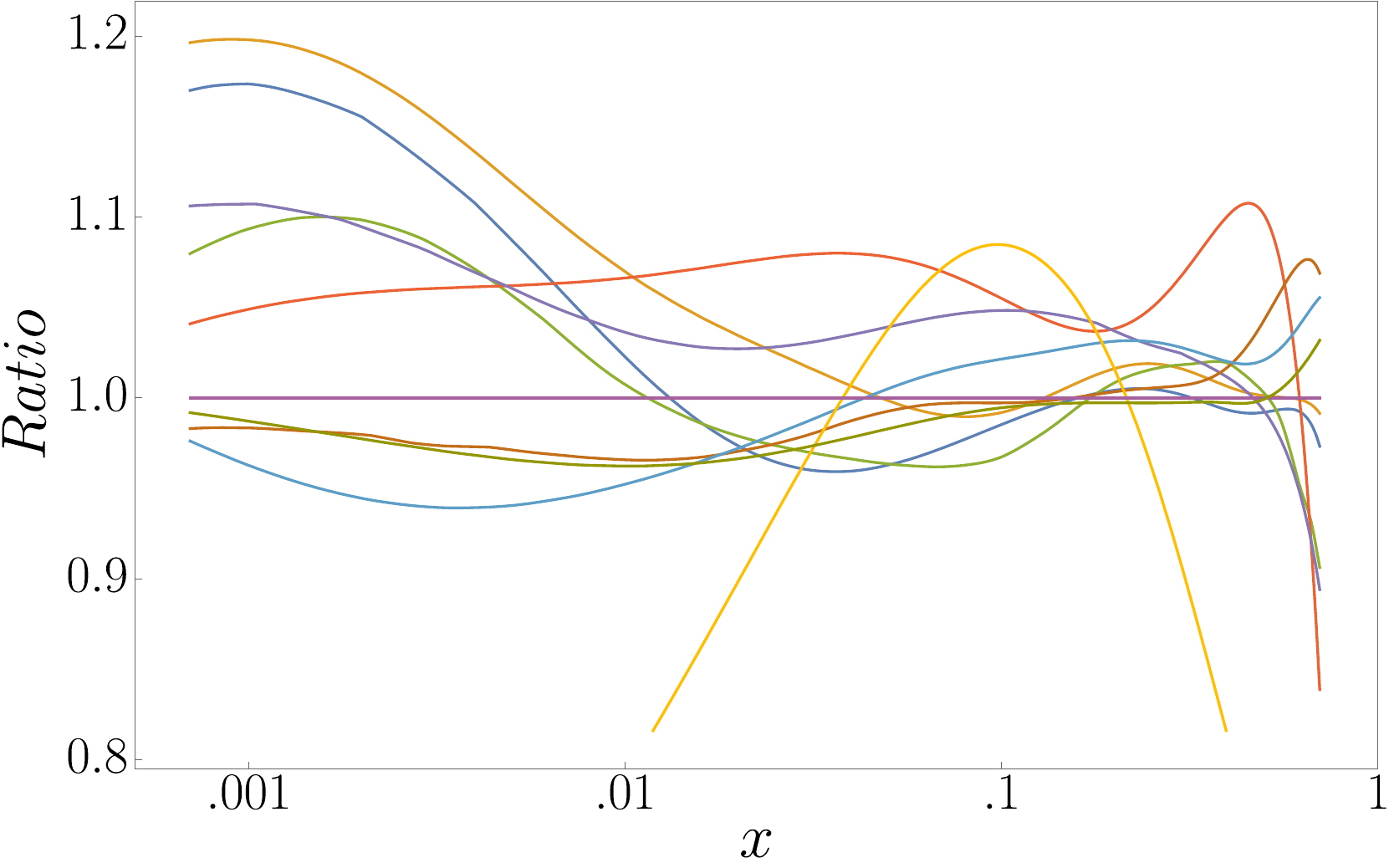}
\protect
\caption{
a)~We display $x\,f_{u}^A(x,Q)$ for the up-quark at $Q = 2~\mathrm{GeV}$ as
a function of $x$ for the 10 PDFs listed in Table~\ref{tab:MomSum}.
\hbox{b)~We display} the ratio of the PDFs in a)~compared to CT10 proton PDF ($A = 1$) 
as a function of x. While we don't identify them individually, the one curve 
(yellow) that distinctly deviates from the others is the nuclear PDF for lead 
$ A = 208 $. 
\label{fig:manyPDFs}}
\end{figure}

Parton Distribution Functions (PDFs) are essential elements for making
predictions involving hadrons (protons and nuclei) in the initial state. 
For example, at the LHC, we can compute the Higgs production cross section 
($\sigma$) using the formula $\sigma_{pp \rightarrow H}=\sum_{a,b} f_{a/P}\otimes f_{b/P}\otimes\omega_{ab\to H}$
where PDFs $f_{a/P}$ and $f_{b/P}$ give the probability density for finding
partons ``$a$'' and ``$b$'' in the two proton beams, and the
hard cross section, $\omega_{ab\to H}$, gives the probability density for partons
$a$ and $b$ producing the Higgs, $H$. The PDFs cannot be computed
from first principles at this time,
so they must be extracted using fits to 
experimental data.\footnote{Lattice QCD has made great strides in 
computing PDFs in recent years.~\cite{Ma:2014jga,Alexandrou:2016tjj}} 
This analysis is performed by a number of collaborations, 
and the PDFs are generally distributed as a grid of values in $x$ and $Q$ 
which must be interpolated to generate the PDF $f_{a/P}(x,Q)$ for flavor ``$a$''
in hadron ``$P$'' at momentum fraction $x$ and energy scale $Q.$

\Mane\footnote{The \Mane\ program was originally developed to run on 
the SMU computing cluster ``ManeFrame'' which is a play on words inspired by 
the school mascot, Peruna the pony.}\ is a flexible, modular, lightweight, 
stand-alone package used to provide access to a wide variety of PDFs within 
\mma. To illustrate the flexibility, in Fig.~\ref{fig:manyPDFs} 
we show how \Mane\ can work simultaneously with different PDF sets from a 
variety of groups.\footnote{All plots presented here have been  generated in \mma.} 
This figure
displays the selected PDF sets listed
in Table~\ref{tab:MomSum}. Some of the sets are in the \lha\ grid 
format\citep{Buckley:2014ana}, and others are in the older PDS grid 
format.\citep{Nadolsky:2008zw} These sets also have different numbers of active 
flavors, $N_{F}$, different values for the initial evolution scale, $Q_{0}$, 
different values for the heavy quark masses, $\{m_{c},m_{b},m_{t}\}$, and they can
represent either free protons or protons bound in nuclei. Nevertheless, \Mane\ is able
to easily compare and contrast sets from different groups in a common
framework. 

As \Mane\ is a stand-alone code, this complements a number of 
other available programs such as 
the QCDNUM program,\cite{Botje:2010ay}
the APFEL program including the web-plotter,\footnote{Details can be 
found in Ref.~\cite{Carrazza:2014gfa} and online at:
{\tt http://apfel.mi.infn.it/}}
the  Transverse Momentum Dependent (TMD) distributions plotter hosted at 
DESY,\footnote{Details can be 
found in Refs.~\cite{Hautmann:2014kza,Hautmann:2014uua}
 and online at:
{\tt http://tmdplotter.desy.de/}}
and also 
the Durham HepData online PDF plotting and calculation 
tool.\footnote{Details can be found online at:
{\tt http://hepdata.cedar.ac.uk/pdf/pdf3.html}}
The online tools provide the ability to quickly plot PDFs, ratios, and luminosities. 
Then with \Mane, it is easy to take the next step and compute
cross sections and other user-selected quantities in the \mma\ environment.

In this paper we describe the key features of \Mane\ available
to the user. 
In Section~\ref{sec:example}, we sketch a minimal
example of how the program is used. 
In Section~\ref{sec:Inside-the-ManeParse}, we provide
some details of how the PDFs are parsed, stored and interpolated. 
In Section~\ref{sec:Sample-Plots}, we display some
example plots that are easily constructed using \Mane. 
In Section~\ref{sec:Error-PDFs:}, we provide examples
of the functions in the \cmd{pdfError} module. 
Finally, 
we discuss files provided by \Mane\ and how to obtain the external
PDF files. 

\section{A simple example\label{sec:example}}

We begin by outlining a simple example of how \Mane\ may be used.
After loading the \Mane\ packages into \mma, the user can
enter the following commands:\\ 
\begin{minipage}{.8\textwidth}
\begin{myquotation}
\cmd{Get[pdfParseLHA.m]}

\cmd{iSet1=pdfParseLHA{[}LHA\_file.info,LHA\_file.dat{]}}

\cmd{pdfFunction{[}iSet1,iParton,x,Q{]}} 

\null 

\cmd{Get[pdfParseCTEQ.m]}

\cmd{iSet2=pdfParseCTEQ{[}PDS\_file.pds{]}}

\cmd{pdfFunction{[}iSet2,iParton,x,Q{]}}

\end{myquotation}
\end{minipage}\\
The first and fourth line load the parsing subpackages included in \Mane.
 Loading either of these, causes the \cmd{pdfCalc} package to be loaded as well.
 The second line reads an \lha\ formatted external data file (\texttt{LHA\_File.dat})
and its associated information file (\texttt{LHA\_File.info}), and generates
an internal PDF set that is referenced by the integer \cmd{iSet1}.
The fifth line reads a PDS formatted external data file\footnote{Note that 
the \lha\ files have both a data file and an info file whereas the older 
\cteq\ PDS files have only a data file. } (\texttt{PDS\_File.pds}) and generates an 
internal PDF set that is referenced by the integer \cmd{iSet2}. 

After reading these data files, the user is provided with the core
function for computing the PDFs: \cmd{pdfFunction{[}iSet,iParton,x,Q{]}.} 
Here, \cmd{iSet} selects the individual PDF set, \cmd{iParton} selects
the parton flavor as shown in Table~\ref{tab:flavTable}, and \{\cmd{x,Q}\} specify
the momentum fraction, $x$, and the energy scale, $Q$, in GeV. 

\cmd{pdfFunction} performs the bulk of the work for the \Mane\
program, so the package has been optimized for speed to make it practical
to perform single or double integrals in a reasonable amount of time;
specifically, the \cmd{pdfFunction} call generally takes
less than 1~ms per core on a standard laptop or desktop. 

Additionally, \Mane\ can handle an arbitrary number of PDF sets
and can switch between sets without delay. When the external PDF file
is parsed, the data is stored internally (about 1~Mb per PDF set) and
the \cmd{iSet} variable essentially functions as a pointer to the
set; thus, it is trivial to loop over many PDF sets as was done in
Fig.~\ref{fig:manyPDFs}. This feature contrasts to some of
the older \fortran\ programs, which could only store a fixed number
of sets in memory and often had to re-read the data files.
 
\begin{table*}[t] 
\renewcommand{\arraystretch}{1.5}
\centering{}%
\begin{adjustbox}{max width=\textwidth}
\begin{tabular}{|c||c|c|c|c|c|c|c|}
\hline
flavor \# & $0$ or $21$ & $\pm 1$ & $\pm 2$ & $\pm 3$ & $\pm 4$ 
  & $\pm 5$ & $\pm 6$   \\
\hline 
parton & gluon & down/dbar & up/ubar & strange/sbar & charm/cbar 
  & bottom/bbar & top/tbar \\
\hline  
\end{tabular}
\end{adjustbox}
\protect
\caption{The standard Monte Carlo (MC) flavor numbering convention\citep{Agashe:2014kda} used within \Mane. 
This differs from the mass-ordered convention used in many older \cteq\ releases. \Mane\ converts these releases into the MC ordering.
\label{tab:flavTable}}
\end{table*}

These are the key elements of the package, however, we also provide many
auxiliary functions described below. Consistent with the \mma\ convention,
all our public functions begin with the prefix ``\cmd{pdf}''.
One can obtain a complete list with the command \cmd{?pdf{*}}.
The usage message for individual functions is displayed in a similar
manner to:

\begin{myquotation}
\cmd{?pdfFunction}
\\[10pt]
pdfFunction[setNumber, flavor, x, Q]
\begin{itemize}
\item This function returns the interpolated value of the PDF for the 
\cteqFile /\dataFile\ file specified by $setNumber$, for the given $flavor$ and 
value of Bjorken $x$ and scale $Q$.
\item \textit{Warning}: The results of this function are only reliable between 
the maximum and minimum values of $x$ and $Q$ in the \cteqFile /\dataFile\ 
file.\footnote{If interpolation outside the given grid is requested by the user, 
\Mane\ is equipped to handle this. 
The \mma\ interpolator will throw a warning message 
and proceed to use built-in extrapolation techniques. 
The \Mane\ interpolator will extrapolate using the behavior defined with \cmd{pdfSetXpower}.}
\end{itemize}
\end{myquotation}
\section{Inside the \Mane\ Package\label{sec:Inside-the-ManeParse}}

\begin{figure*}[t] 
\centering{}
\includegraphics[width=0.8\textwidth]{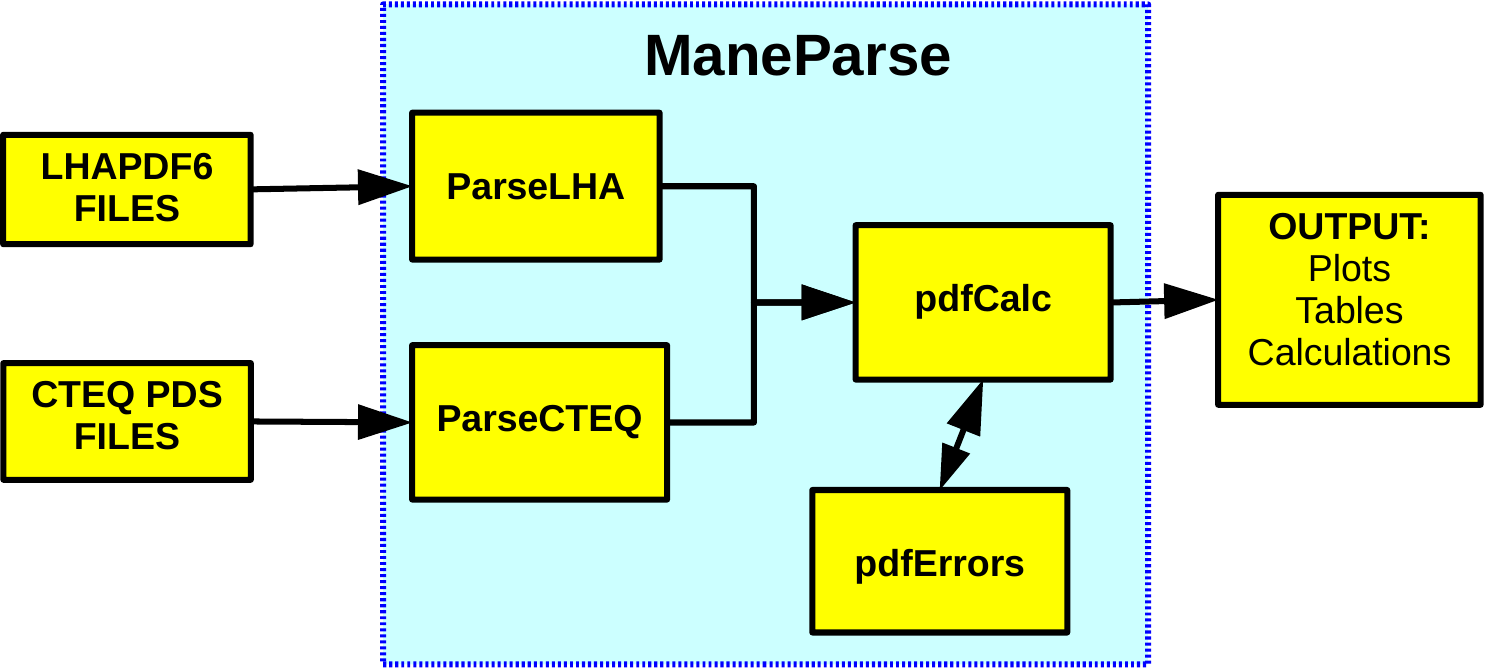}
\protect
\caption{A schematic overview of the \Mane\ package and the individual modules.
\label{fig:Flow-chart:}}
\end{figure*}

\subsection{Overview of package}

\Mane\ internally consists of four modules (or sub-packages) as illustrated in 
Fig.~\ref{fig:Flow-chart:}. The modular structure of \Mane\ allows for separate 
parsers for the \lha\ (\cmd{pdfParseLHA}) and PDS (\cmd{pdfParseCTEQ}) 
grids which read the individual file types and pass the information on to a 
common calculation (\cmd{pdfCalc}) module. 

The new \lha\ format is intended as a standard that all groups can
use to release their results. Additionally, many older PDF sets have
been converted into this format.

The \Mane\ modular structure provides flexibility,
as the user can use both \lha\ and PDS format, or even write a custom
parser to read a set that is not in one of these formats.

The error PDFs module (\cmd{pdfError}) uses \cmd{pdfCalc} to construct 
PDF uncertainties, luminosities, and correlations as illustrated in 
Sec.~\ref{sec:Error-PDFs:}.

The key elements of each PDF set include the 3-dimensional $\{x,Q,N_{F}\}$
grid and the associated information, which is stored as a set of \mma\
rules. We now describe the features and some details of these structures.

\subsection{The PDF \texorpdfstring{$\{x,Q,N_{F}\}$}{\{x,Q,NF\}} grid}

The parsing routines \cmd{pdfParseLHA}
and \cmd{pdfParseCTEQ} read the external files and assemble
the PDF sets into a common data structure that is used by the \cmd{pdfCalc}
module. The central structure is a 3-dimensional grid of PDF values
in \{x,Q,$N_{F}$\} space, which uses vectors $\{x_{vec},Q_{vec}\}$
to specify the grid points. The spacing of $\{x_{vec},Q_{vec}\}$
need not be uniform; typically, $Q_{vec}$ uses logarithmic spacing,
and $x_{vec}$ is commonly logarithmic at small $x$ and linear at
large $x$. Different spacings in $x_{vec}$ and $Q_{vec}$ do not pose a problem 
for the \cmd{pdfCalc} package, as the grid points are simply interpolated to 
provide the PDF at a particular point in \{x,Q,$N_F$\}. The user is agnostic to 
the specific grid spacing chosen in a PDF release.

\subsubsection{\texorpdfstring{$N_{F}$}{NF} Convention}

The $N_{F}$ flavor dimension is determined by the \cmd{iSet} value
passed to \cmd{pdfFunction}. The association between the grid slice in 
$N_{F}$ and \cmd{iSet} is specified in the \lha\ info file using the 
``\cmd{key:data}'' format such as ``\cmd{Flavors: $[-5,-4,-3,-2,-1,1,2,3,4,5,21]$}''. This 
tells us which partons are in the grid, and their proper order.\footnote{For 
the PDS files, this information is contained in the header of the data file so 
there is not a separate info file; \cmd{pdfParseCTEQ} extracts the proper 
association.} Note: we use the standard Monte Carlo (MC) convention\footnote{See Ref.~\citep{Agashe:2014kda} 
``Review of Particle Physics,'' Chapter~34 entitled ``Monte Carlo particle 
numbering scheme.''} throughout \Mane\ where $d=1$ and $u=2$ rather than the mass-ordered 
convention (see Table~\ref{tab:flavTable}).\footnote{Caution is required here as many of the older \cteq\ 
releases use the mass-ordered convention with $u=1$ and $d=2$. \Mane\ converts 
these mass-ordered sets into the MC ordering.} The standard 
MC convention also labels the gluon as \cmd{iParton}\ $=21$;
for compatibility, the gluon in \Mane\ can be identified with either
\cmd{iParton}\ $=21$ or \cmd{iParton}\ $=0$.

\Mane\ is able to work with PDF sets with different numbers of
flavors. For example, in Fig.~\ref{fig:manyPDFs}, the \nnpdf\ set includes
$N_{F}= 6$ where \cmd{iParton} $ = \{\bar{t},\ldots ,t\}$, while most of the 
other sets have $N_{F}=5$. If a flavor, \cmd{iParton}, is not defined, 
\cmd{pdfFunction} will return zero. This feature allows the user to write 
a sum over all quarks $\sum\,f_{i}(x,Q)$ for $i=\{-6, \ldots ,6\}$ without worrying
whether some PDF sets might have less than 6 active flavors. 

Additionally, the \Mane\ framework has the flexibility to handle
new particles such as a 4\textsuperscript{th} generation of quarks with 
\cmd{iParton} $=\{b',t'\}=\{7,8\}$ or a light gluino with 
\cmd{iParton} $=\tilde{g}=1000021$ PDF by identifying the flavor index, 
\cmd{iParton}, with the appropriate grid position in the \lha\ info file.

\subsubsection{Q Sub-Grids}

\begin{figure*}[t] 
\centering{}
\includegraphics[width=0.45\textwidth]{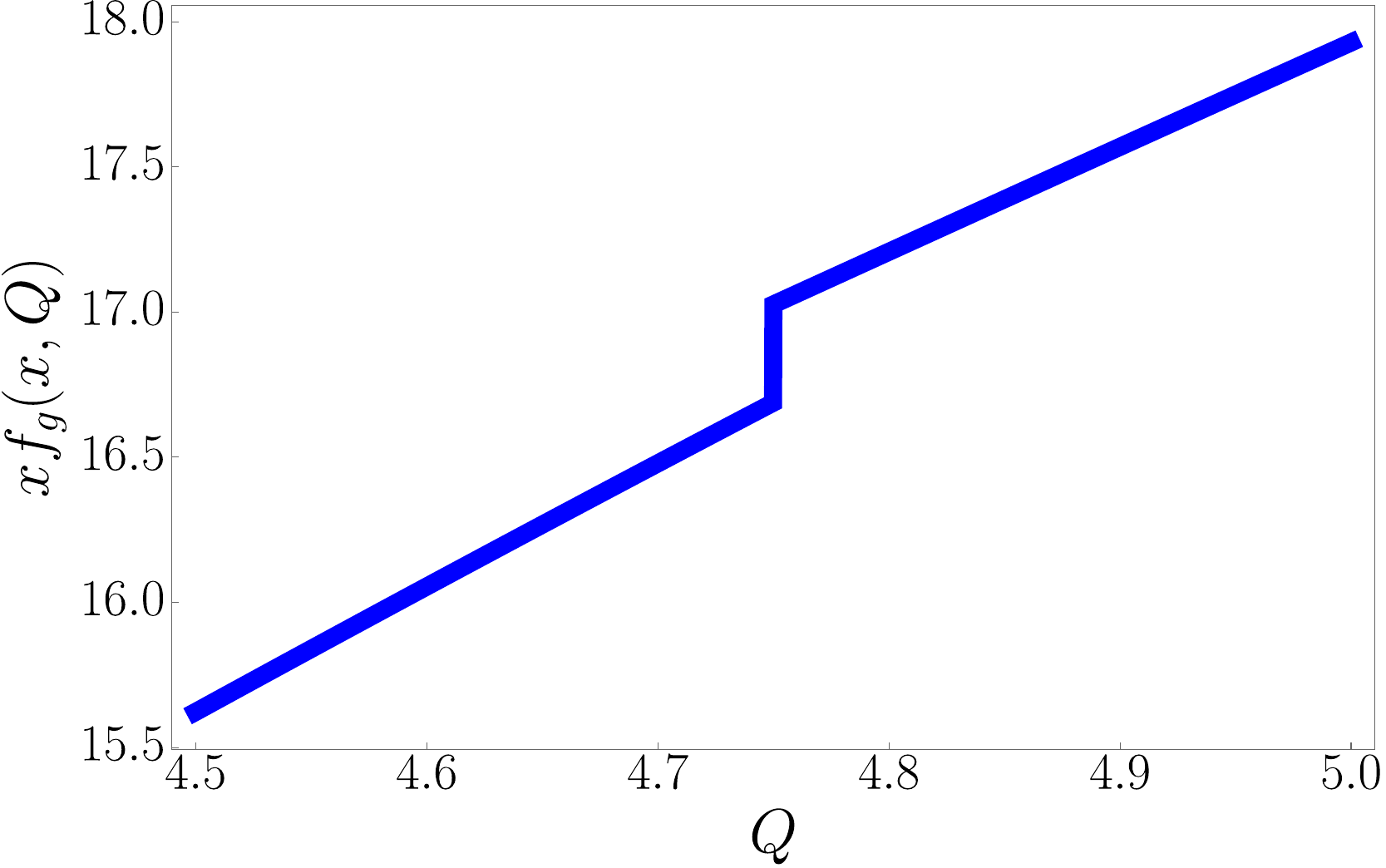}
\hfil
\includegraphics[width=0.45\textwidth]{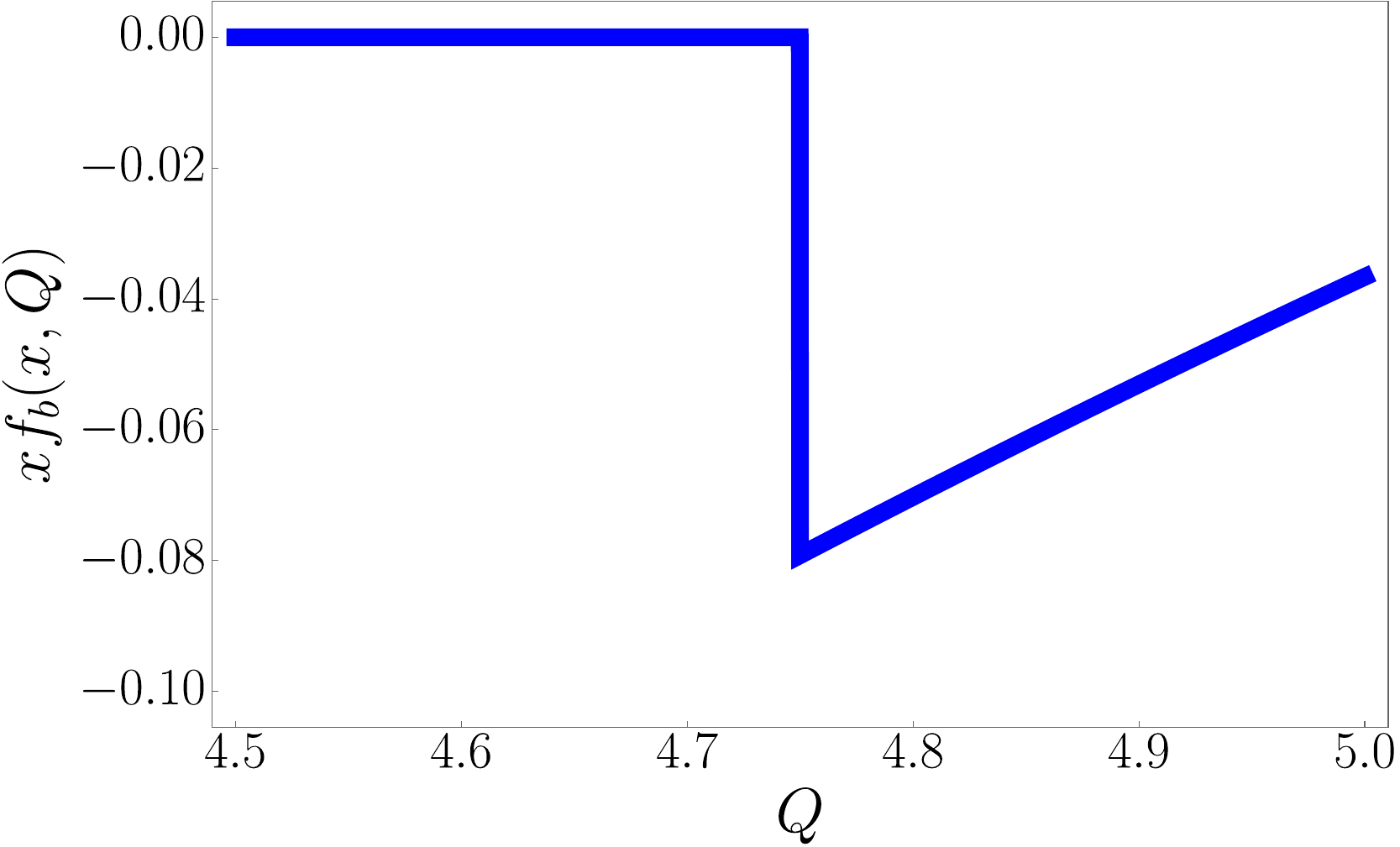}
\protect
\caption{The discontinuity of the gluon (left) and b-quark (right) PDFs across
the $m_{b}=4.75\ \mathrm{GeV}$ flavor threshold; the horizontal axis is $Q$ (in GeV),
and the vertical axis is $x\,f(x,Q)$. The curves are for the MSTW2008nnlo68cl
PDF with $x=10^{-4}$. Note that the gluon and b-quark shift in opposite
directions to ensure the momentum sum rule is satisfied. 
\label{fig:mstw-disc }}
\end{figure*}

\begin{figure*}[t] 
\centering{}
\includegraphics[width=0.45\textwidth]{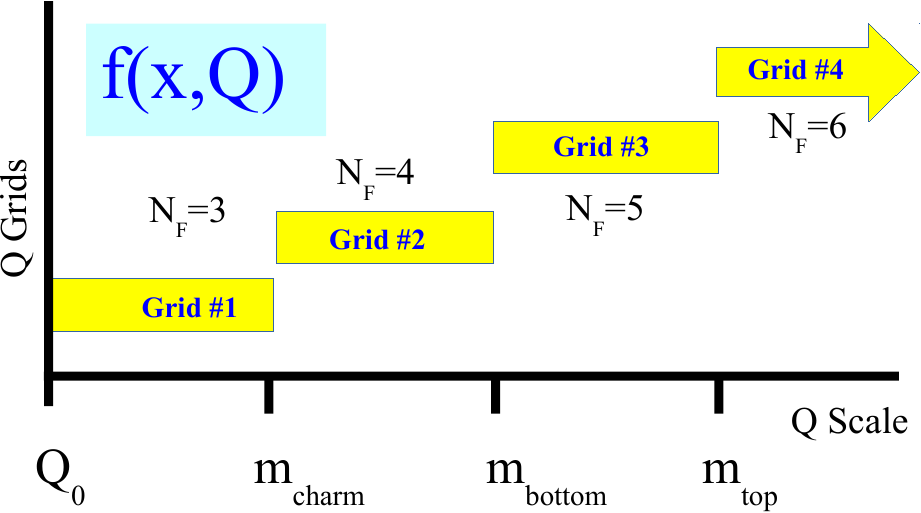}
\hfil
\includegraphics[width=0.45\textwidth]{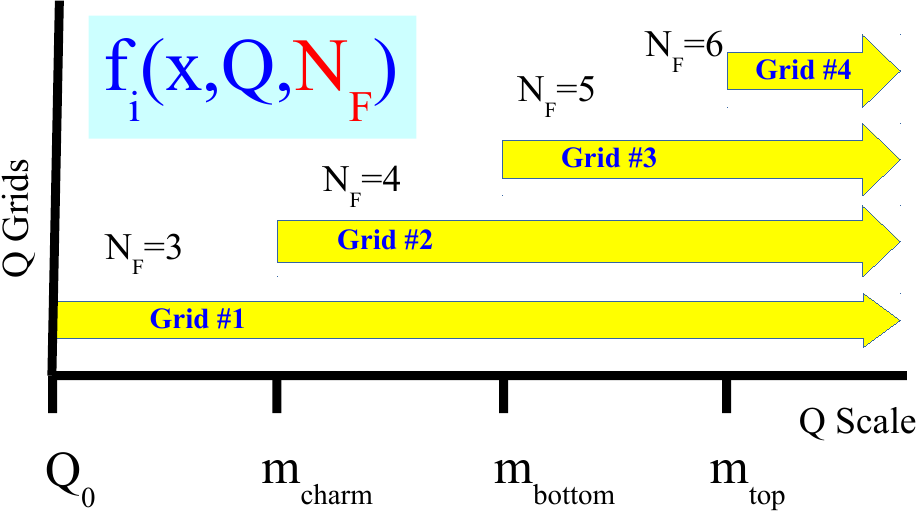}
\protect
\caption{A schematic representation of the $Q$ sub-grids used to handle 
discontinuities across the heavy quark thresholds at $\{m_{c},m_{b},m_{t}\}$. 
Fig.~a) shows the conventional arrangement for $f(x,Q)$ with non-overlapping
sub-grids; for a given $Q$, the $N_{F}$ flavor dimension is uniquely determined.
Fig.~b) shows a flavor-number-dependent PDF $f(x,Q,N_{F})$ where the user has
the freedom to choose the $N_{F}$ flavor dimension value (and hence the 
sub-grid).
\label{fig:qGrids}}
\end{figure*}

At NNLO and beyond, the PDFs can become discontinuous across
the mass flavor thresholds. This is illustrated using the NNLO MSTW
set in Fig.~\ref{fig:mstw-disc } where we observe a discontinuity
of both the gluon and b-quark PDF across the b-quark threshold at 
$m_{b}=4.75$~GeV. \Mane\ accommodates this by using sub-grids in $Q$ as 
illustrated in Fig.~\ref{fig:qGrids}-a); for example, we use separate grids
below and above the threshold at $Q=m_{b}=4.75\ \mathrm{GeV}$. When we call
the PDF at a specific $Q$ value, \Mane\ looks up the relevant
heavy quark thresholds, $\{m_{c},m_{b},m_{t}\}$, to determine which
sub-grid to use for the interpolation. For $Q < m_{b}$, sub-grid \#2
($N_{F}=4$) is used, and for $Q \geq m_{b}$, sub-grid \#3 ($N_{F}=5$)
is chosen. 

Note that for the $x$ value ($10^{-4}$) displayed in Fig.~\ref{fig:mstw-disc }, 
the b-quark PDF is negative for $Q$ just above $m_{b}$; this is
the correct higher-order result and justifies (in part) why we do
not force the PDFs to be positive definite. This behavior also makes
sense in terms of the momentum sum rule, which we will discuss in
Sec.~\ref{sub:Momentum-Sum-Rules}.

\subsubsection{An \texorpdfstring{$N_{F}$}{NF}-dependent PDF: 
 \texorpdfstring{$f(x,Q,N_{F})$}{f(x,Q,NF)} }

Note, the use of sub-grids in $Q$ also enables the use of 
overlapping $N_{F}$ ranges as in a hybrid scheme as described in
Ref.~\citep{Kusina:2013slm}; in this case, we generalize the PDF
so that it also becomes a function of the number of flavors: 
$f(x,Q,N_{F})$. This feature is useful if, for example, we are 
performing a fit to data in the region $Q\sim m_{b}$; we can perform a 
consistent $N_{F}=4$ flavor fit even if some of the data are above the 
$N_{f}=5$ threshold ($Q>m_{b}$) by selecting $f(x,Q,N_{F}=4)$; thus, we avoid 
encountering any discontinuities in the region of the data.\footnote{Note 
that the APFEL PDF evolution library\cite{Bertone:2013vaa} is in the 
process of implementing these features.}
We illustrate this
generalized case for $f(x,Q,N_{F})$ in Fig.~\ref{fig:qGrids}-b).
Here, the user has the freedom to choose the active number of flavors,
$N_{F}$, rather than being forced to transition at the quark mass
values as in Fig.~\ref{fig:qGrids}-a). 

\subsection{The \lha\ Info File}

\begin{table*}[t]  
\renewcommand{\arraystretch}{1.5}
\begin{adjustbox}{max width=\textwidth}
\centering{}%
\begin{tabular}{|c|c|}
\hline 
YAML  & \mma\tabularnewline
\hline 
\hline 
key: ``data''  & ``key'' $\rightarrow$ ``data''\tabularnewline
\hline 
SetDesc: ``nCTEQ15 ...''  & ``SetDesc'' $\rightarrow$ ``nCTEQ15 ...''\tabularnewline
\hline 
NumFlavors: 5  & ``NumFlavors'' $\rightarrow$ $5$\tabularnewline
\hline 
Flavors: {[}-5,-4,-3,-2,-1,1,2,3,4,5,21{]} & ``Flavors''$\rightarrow$\{-5,-4,-3,-2,-1,1,2,3,4,5,21\}\tabularnewline
\hline 
AlphaS\_Qs: {[}1.299999e+00, ...{]}  & ``AlphaS\_Qs''$\rightarrow \{ 1.299999\times10^{+00}, ... \}$\tabularnewline
\hline 
UnknownKey: data  & ``UnknownKey''$\rightarrow$''data''\tabularnewline
\hline 
\end{tabular}
\end{adjustbox}
\protect
\caption{Sample YAML entries contained in the \lha\ info file, and the corresponding
rules passed to \mma. The rules for a specific PDF set are
obtained using the \cmd{pdfGetInfo{[}iSet{]}}function. 
\label{tab:yamlTable}}
\end{table*}

In addition to the 3-dimensional $\{x,Q,N_{F}\}$ grid, there is auxiliary
material associated with each PDF set. In the LHA format, each PDF
collection has an associated ``info'' file which contains the additional
data in a YAML format,\footnote{``YAML Ain't Markup Language'' 
\url{http://yaml.org/} } whereas in the CTEQ PDS format files, the auxiliary information is contained
at the top of each PDS data file. Each parser interprets this information
and builds a list of \mma\ rules.

The basic syntax of YAML is \cmd{{[}key: ``data''{]}}, and the LHA parser
converts this into a \mma\ rule as 
\cmd{\{``key''$\rightarrow$``data''\}}. This can be viewed within \Mane\ using the 
function \cmd{pdfGetInfo{[}iSet{]}}, and Table~
\ref{tab:yamlTable} demonstrates  some sample mappings between the two.

If ``\cmd{key}'' is known to be a number, ``\cmd{data}'' is converted 
from a string into a number. This behavior applies to values such as 
\{NumFlavors, QMin, MTop, ...\}. 
If ``\cmd{key}'' is known to be a list such as \{Flavors, AlphaS\_Qs\},
``\cmd{data}'' is converted from a string into a \mma\ list.
If ``\text{key}'' is unknown, ``\cmd{data}'' is left as a string.
This means that \Mane\ can handle any unknown ``\cmd{key}'', and the
user can modify these rules after the fact, or introduce a custom
modification by identifying ``\cmd{key}'' to the parser.

\subsection{Interpolation}

Once the 3-dimensional \{x,Q,$N_{F}$\} grid and auxiliary rules are
given to the \cmd{pdfCalc} module, we are ready to interact with the PDFs. When the
user calls for $f_{i}(x,Q)$, the \cmd{pdfCalc} module will determine the
appropriate $N_{F}$ index and $Q$ grid and do a 4-point interpolation
in the 2-dimensional $\{x,Q\}$ space. For the interpolation we use a 4-point Lagrange interpolation given by:\footnote{We present the interpolation formulas in the $x$-variable; an equivalent
form is used for the $Q$ interpolation.}

\[
g(x)=c_{0}(x)\,y_{0}+c_{1}(x)\,y_{1}+c_{2}(x)\,y_{2}+c_{3}(x)\,y_{3}
\]
where $y_{k}=g(x_{k})$ are the PDF values at the grid points, and
the coefficients are given by: 

\[
c_{j}(x)=\sideset{}{'}\prod_{0\leq m\leq3}\;\frac{(x-x_{m})}{(x_{j}-x_{m})}
\]
where the prime $(')$ indicates the restriction $j\not=m$ in the
product. This formula has the feature that the interpolated curve
will always contain the grid points $\{x_{i},y_{i}\}$. The grid points
do not need to be equally spaced. 

To perform the 2-dimensional interpolation, we extract a $4\times4$
sub-grid in $\{x,Q\}$ space; we first compute 4 interpolations in
$x$-space, and then use these to perform a 4-point interpolation
in $Q$-space. Generally, \cmd{pdfCalc} will interpolate
$\{x,Q\}$ values with 2 grid points on each side, but at the edges
of the grid, it will use a 3-1 split. It also will extrapolate beyond
the limits of the grid and will return a number, even if it is unphysical.
Except for setting $f_{i}(x,Q)=0$ for $x>1$, we do not check bounds,
as this would slow the computation; in the sample files, we do provide
examples of how the user can implement particular boundaries if desired.

Additionally, we allow the interpolated PDF to be negative. At very
large $x$ this can happen due to numerical uncertainty, but there
are also instances where a negative PDF is the physical result, such
as at NNLO (illustrated in Fig.~\ref{fig:mstw-disc }). Within \mma,
it is easy for the user to impose particular limits (\textit{i.e.}
positivity) if desired.
The interpolation can be performed either with the \mma\ \cmd{Interpolate}
function (default) or a custom 4-point Lagrange interpolator and is set with the 
\cmd{pdfSetInterpolator} function. We set
the \mma\ \cmd{Interpolate} function as the default, as it is slightly
faster, but the custom 4-point Lagrange interpolator often will provide
better extrapolation of the PDFs beyond the grid boundaries and has
some adjustable parameters which are useful in the small $x$ region. 

The PDF typically increases as $1/x^{a}$ at small $x$ where $a\sim1.5$;
thus, we can improve the interpolation by scaling the PDF by a factor
of $x^{a}$ which is implemented by replacing $y_{k}\to x^{a}\,g(x_{k})$
and $g(x)\to g(x)/x^{a}$. This is why many of the PDF programs fundamentally
compute with $xf(x)$ rather than $f(x)$. To return $f_{i}(x,Q)$ we divide 
by $x$, but to avoid dividing by zero we internally impose a default 
minimum $x$ value of $x_{min}=10^{-30}$. The default scaling factor 
for the the custom interpolator is $a=1$, but this can be set with the 
\cmd{pdfSetXpower{[}a{]}} function. 

\subsubsection{Interpolation Quality}

\begin{figure*}
\begin{subfigure}{.5\textwidth}
  \centering
  \includegraphics[width=.8\linewidth]{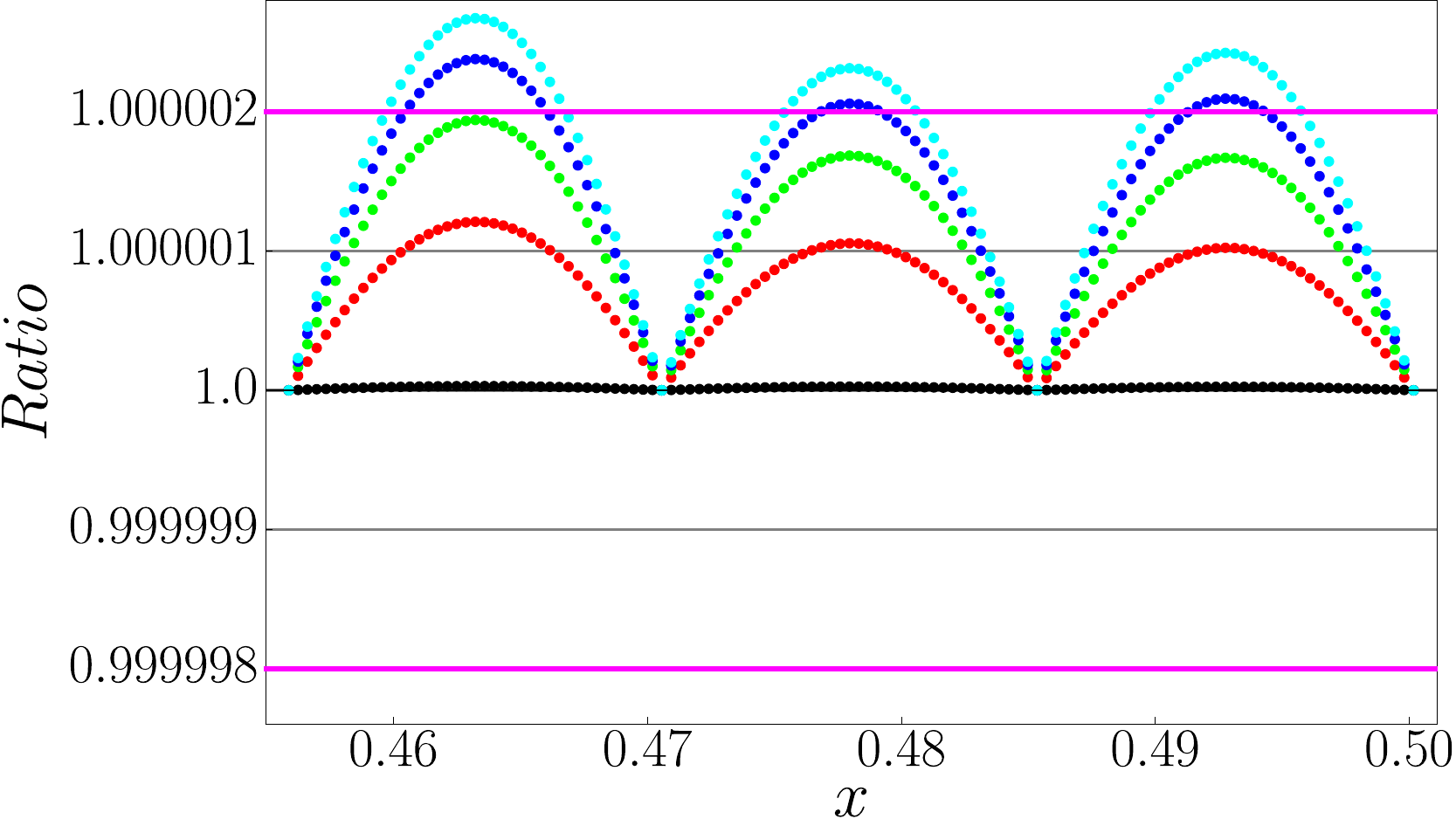}
  \caption{}
  \label{fig:interp_highX}
\end{subfigure}%
\begin{subfigure}{.5\textwidth}
  \centering
  \includegraphics[width=.8\linewidth]{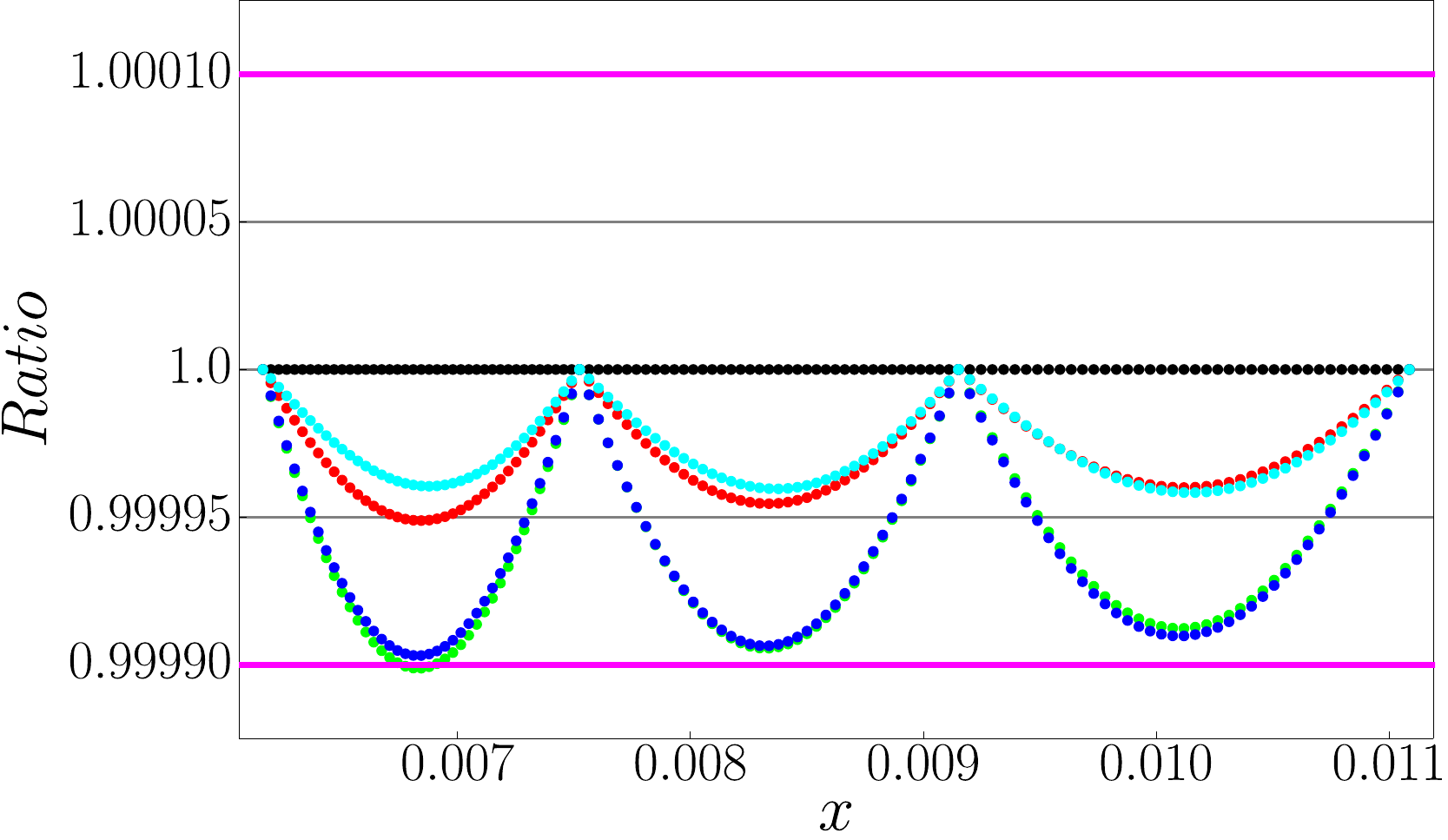}
  \caption{}
  \label{fig:interp_midX}
\end{subfigure}\\
\begin{subfigure}{.5\textwidth}
  \centering
  \includegraphics[width=.8\linewidth]{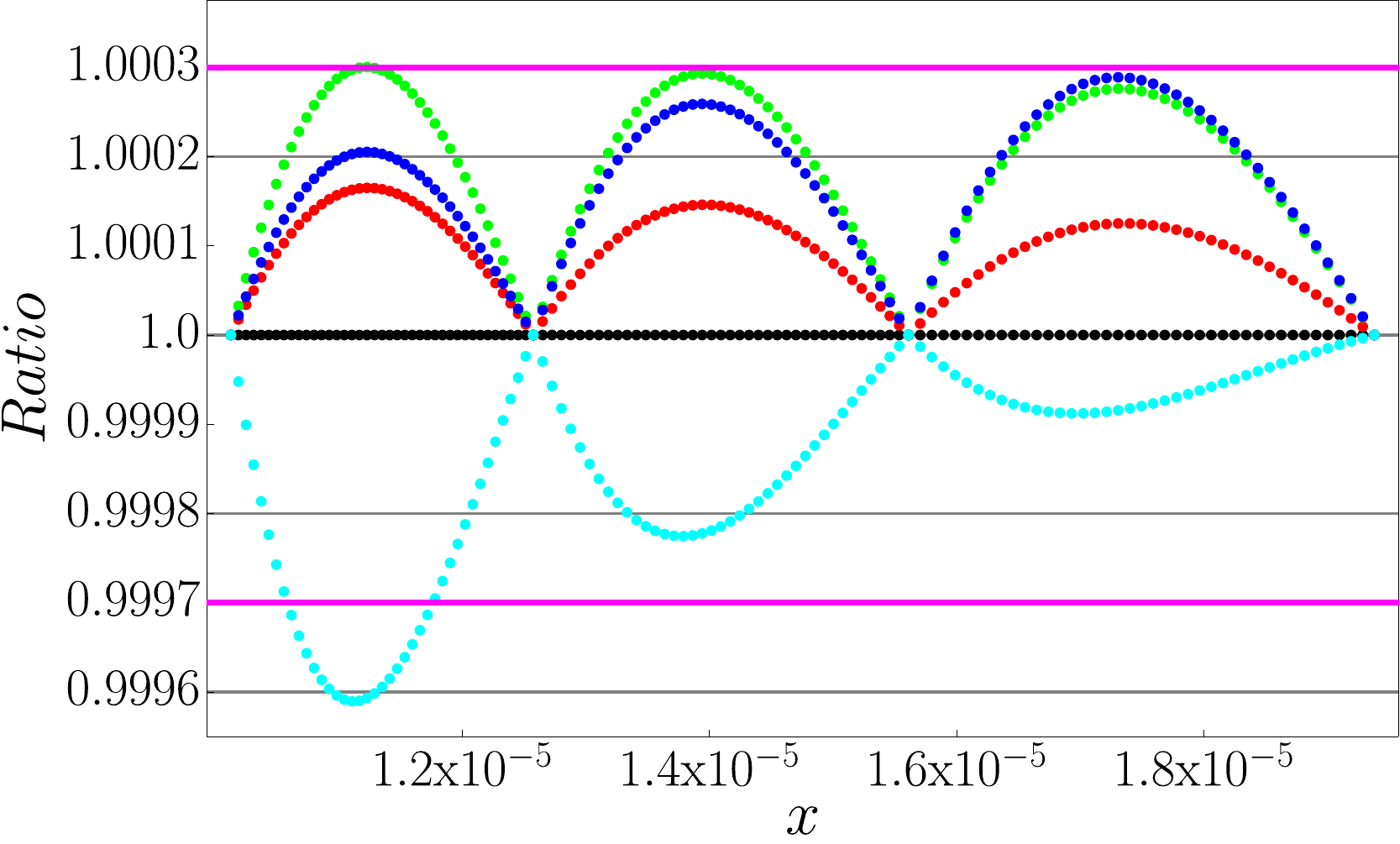}
  \caption{}
  \label{fig:interp_lowX}
\end{subfigure}%
\begin{subfigure}{.6\textwidth}
  \centering
  \includegraphics[width=.6\linewidth]{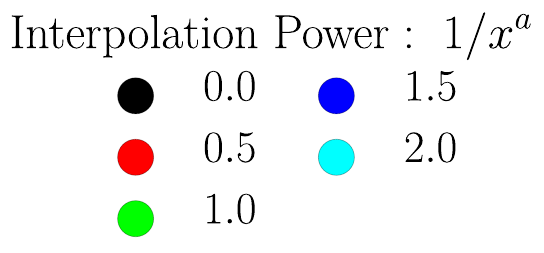}
  \label{fig:interp_legend}
\end{subfigure}\\
\begin{subfigure}{.5\textwidth}
  \centering
  \includegraphics[width=.8\linewidth]{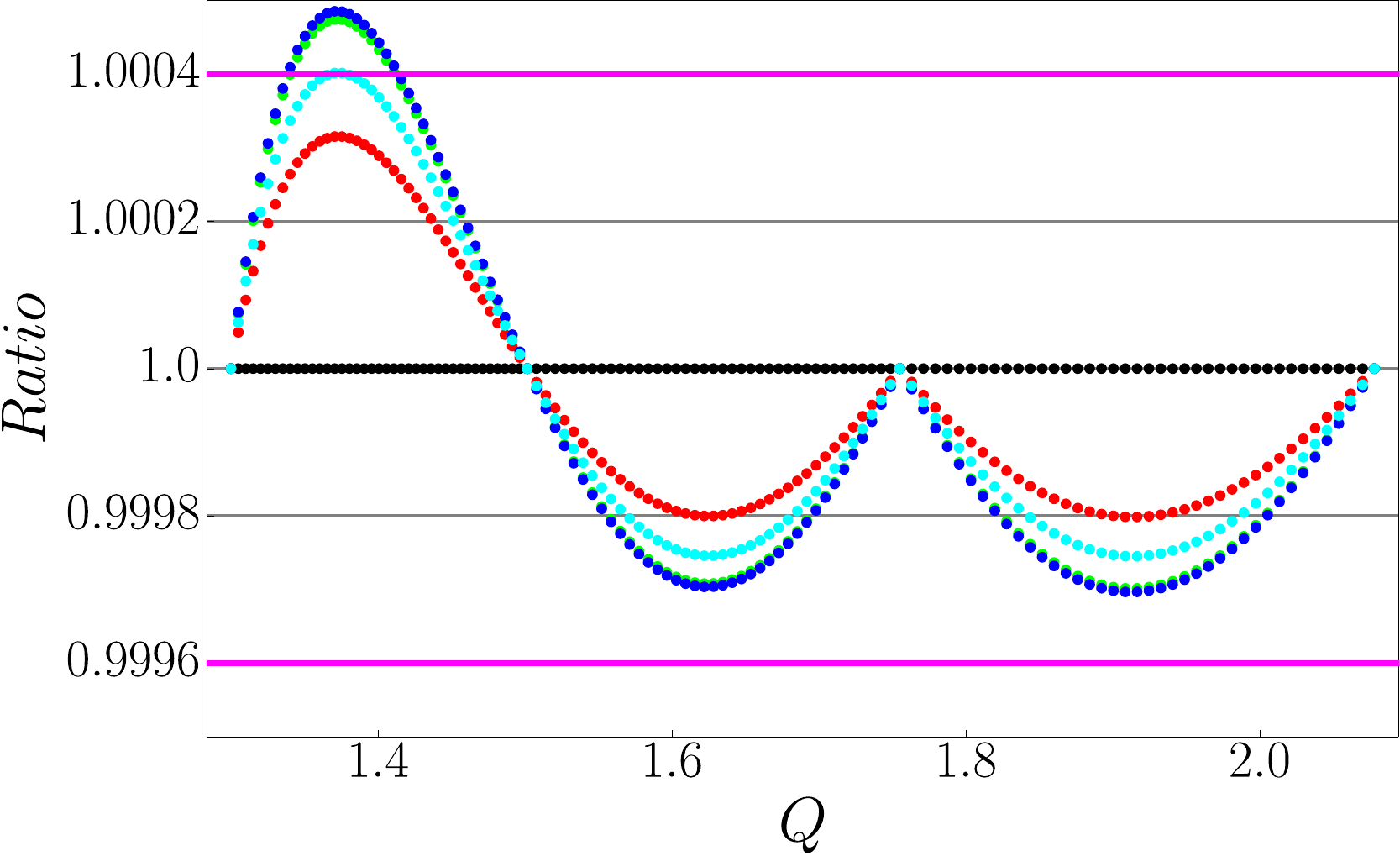}
  \caption{}
  \label{fig:interp_lowQ}
\end{subfigure}%
\begin{subfigure}{.5\textwidth}
  \centering
  \includegraphics[width=.8\linewidth]{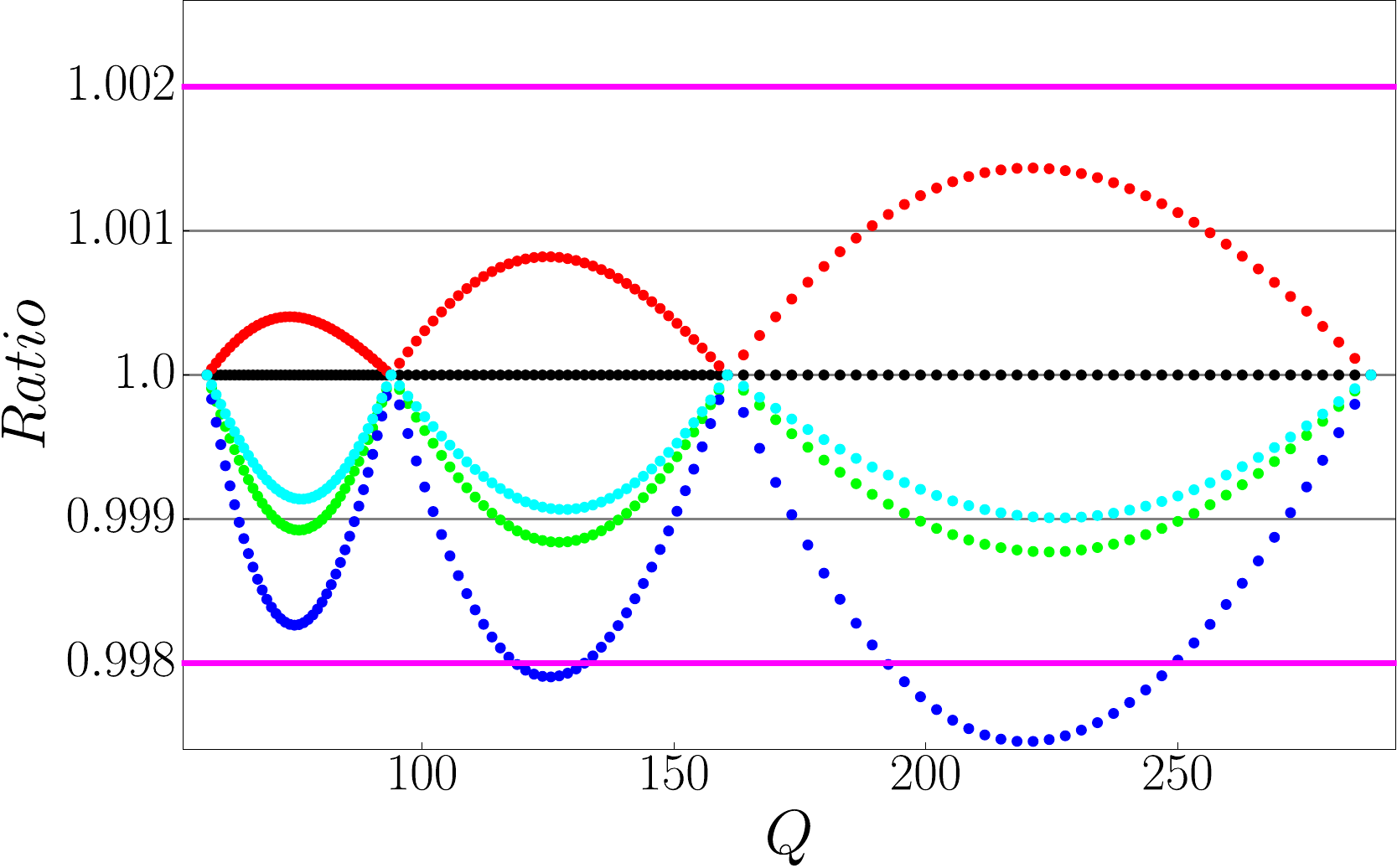}
  \caption{}
  \label{fig:interp_midQ}
\end{subfigure}\\
\caption{We show the numerical variation of the interpolation for the CT10
central set by presenting the ratio of the \Mane\ interpolator with different $a$ values 
to the \mma\ interpolator. 
The range of the CT10 grid is $x=[10^{-8},1]$ with 150 points, 
and $Q=[1.3,\,34515]\,GeV$ with 24 points.
In (a)-(c), we display the variation in $x$ for fixed
$Q=1.3\,GeV$ (which is a grid value).
In (d)-(e), we display the variation in $Q$ for fixed
$x=0.0110878$ (which is a grid value).
We have drawn horizontal guide-lines to indicate the approximate numerical
variation. 
The ratios were plotted as points rather than lines to avoid any line-smoothing
of the graphics output.
}
\label{fig:interp}
\end{figure*}

By construction, the interpolation curve will always intersect the
grid values: $g(x_{i},Q_{j})\equiv f(x_{i},Q_{j})$ if $x_{i}$ and
$Q_{j}$ are grid points. Therefore, the numerical uncertainty arises
from how we connect these grid points.
We have bench-marked many of the PDF sets to ensure our interpolations
are accurate across the defined grid in $\{x,Q\}$ space. For the
PDS files, our interpolation (with scaling $1/x^a$ for $a=1.0$)  uses the
same algorithm as the benchmark \cteq\ \fortran,  so our results easily 
match to better than one part in $10^{3}$.
The \lha\ interpolation uses a logarithmic bi-cubic interpolation in the central region, 
and switches to linear near the grid 
boundaries.\footnote{\lha\ has validated a number of PDF sets, 
and these generally match both our interpolator, with $a=1$, and the \mma\ 
interpolator to 1 part in $10^{-3}$.}
To illustrate the range of
numerical uncertainty, we will show how the interpolation changes
as we vary the $a$ power. We will also compare with the built-in
\mma\ interpolator. 
If a different interpolation  is required,  the $a$-parameter can be tuned, 
or the user can supply a custom interpolation routine.

In Fig.~\ref{fig:interp}, we show the ratio of the interpolated value for the
gluon PDF compared to the default \mma\ interpolation. We select
a Q value which is precisely a grid point, and then show the variation
as a function of x between these grid values. Figs.~\ref{fig:interp_highX},
\ref{fig:interp_midX},\ref{fig:interp_lowX} show the results for three ranges
of $x$, \{small, mid, large\}, while Figs.~\ref{fig:interp_lowQ},\ref{fig:interp_midQ} 
show the results for small $Q$ and mid $Q$.
In all five plots, we observe that the interpolated curves match exactly
at the grid values ($x_{k}$), as they should. In between
the grid values, we see there is a variation depending on the details
of the interpolation and the particular value of the scaling power
$a$. We have varied the scaling power over the range $a=$\{0.0,0.5,1.0,1.5,2.0\}.
The scaling power $a=0$ matches with the default \mma\ interpolation
routine, while $a=1$ compensates for the $1/x$ PDF behavior
at small $x$.

In Fig.~\ref{fig:interp_highX},  we observe that the variation is quite small
in the large $x$ range ($x\sim0.5$) , of order $\sim3\times10^{-6}$. 
For many calculations, such as Higgs and $W/Z$ boson production, 
the mid $x$ range of ($x\sim0.01$), seen in Fig.~\ref{fig:interp_midX},
 is the most relevant region and here 
we find the variation to be a bit larger, of order $\sim1\times10^{-4}$.
At the small $x$ range ($x\sim10^{-5}$), Fig.~\ref{fig:interp_lowX}, 
we find the largest variation which can be of order
$\lesssim10^{-3}$; this is partly because the PDFs are diverging
in the limit $x\to0$, so the relative error
 increases.\footnote{The PDFs typically exhibit a rise at small $x$ of the form $1/x^{a}$.
At smaller $Q$ values, the exponent is commonly slightly larger than
$1$, and increases with increasing $Q$ toward an asymptotic limit
in the range $a\sim[1.5,1.7]$. Note, the momentum sum rule requires
$a<2$.\cite{Lomatch:1988uc}}

We now investigate the quality of the interpolation in the $Q$ variable. 
In ~\ref{fig:interp_lowQ}, we show the small $Q$ range,
($Q\sim Q_{0}$). Here, the steps in Q are about 20\% apart and we
see the variation is of order $\sim5\times10^{-4}$. At the larger $Q$ range
in ~\ref{fig:interp_midQ}, the steps in $Q$ are up to 100\% apart and we see the variation
is of order $\sim10^{-3}$; if increased accuracy is required here,
the obvious solution would be to include more gird points in $Q$. 

In general, we expect $a=1$ yields the best representation of the PDFs, and 
the spread between $a=0$ and $a=1$ is a reasonable estimate of the uncertainty. 
Computing  the momentum sum rule ({\it c.f.}, Table~\ref{tab:MomSum}) can also provide a useful check.

\begin{figure}[t] 
\centering{}
\includegraphics[width=0.45\textwidth]{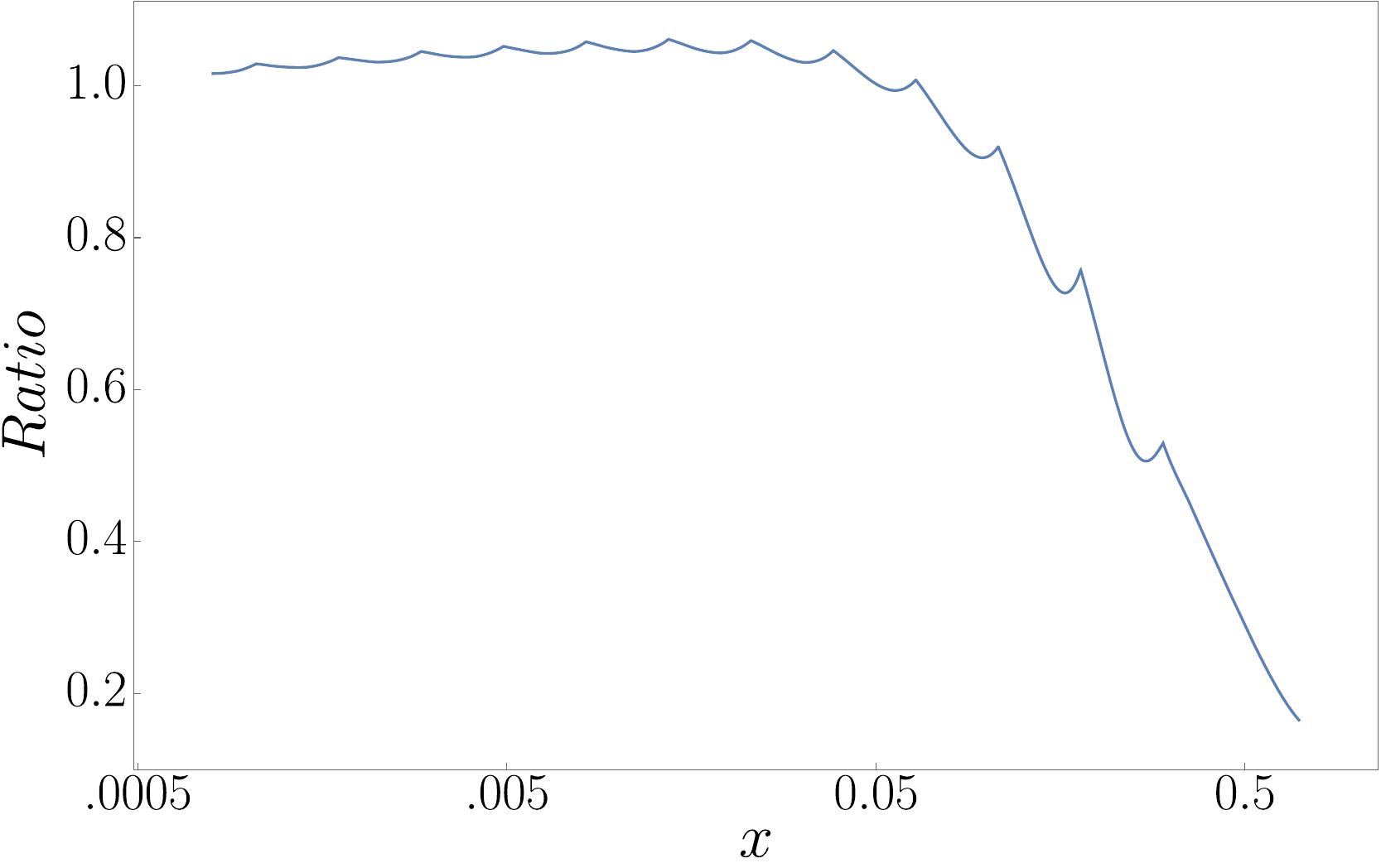}
\hfil
\includegraphics[width=0.45\textwidth]{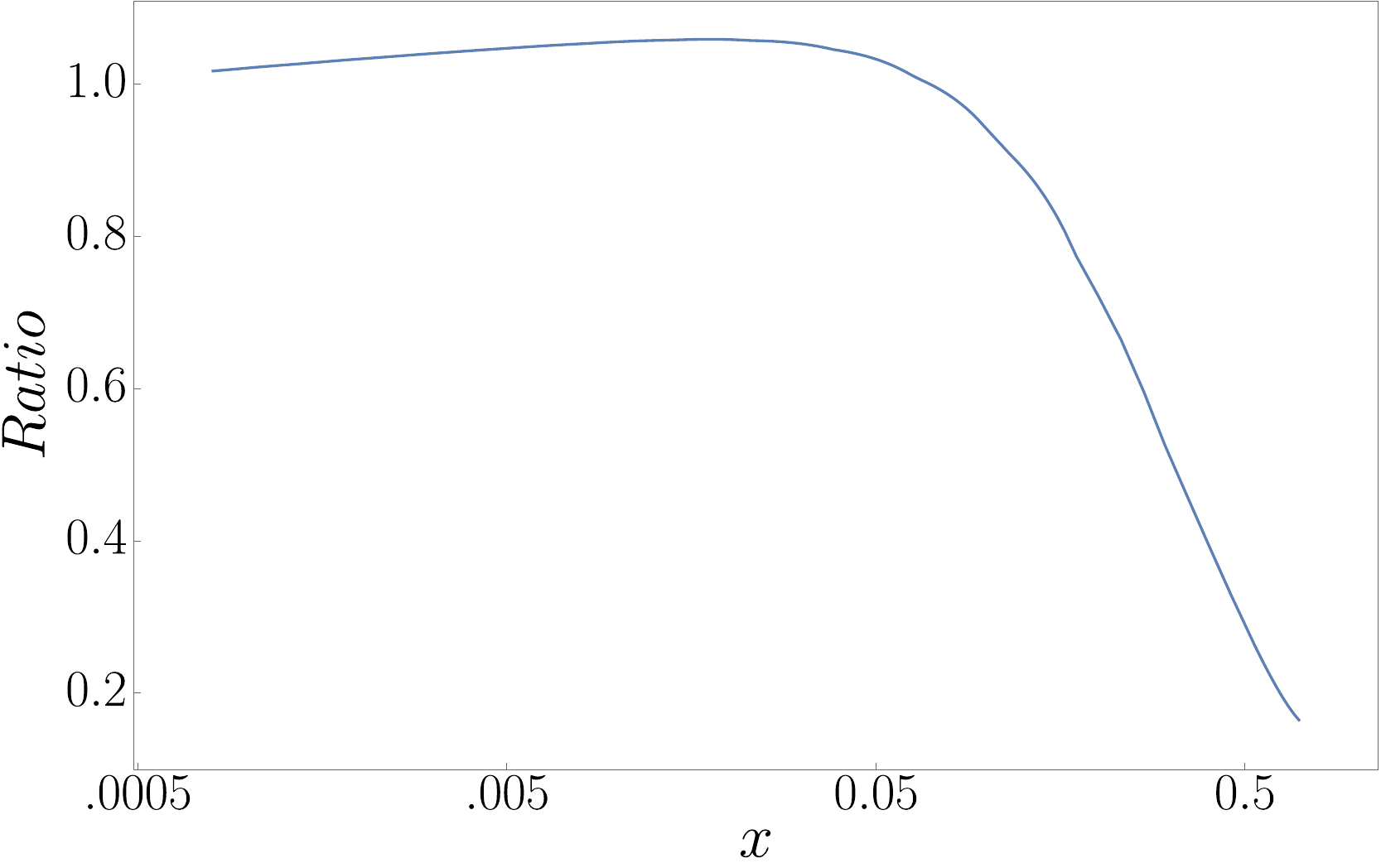}
\protect
\caption{The ratio of PDFs can sometimes lead to interpolation problems; we
display the ratio of two gluon PDFs at $Q = 100\ \mathrm{GeV}$. Fig.-a) on
the left was generated with the default \mma\ interpolator,
and Fig.-b) on the right was generated with the custom 4-point Lagrange
interpolation with the default scaling of $a = 1$. 
\label{fig:badInterp}} 
\end{figure}

We find that ratios of PDFs are more sensitive to the interpolation than the PDFs themselves.
For illustrative purposes, in Fig.~\ref{fig:badInterp}, we show an
example of a poor interpolation generated with the \mma\ interpolator
compared to a good interpolation by the custom 4-point Lagrange
interpolation with the default $a = 1$ scaling; in general, we find
the custom 4-point Lagrange interpolation computes smoother ratios
and provides better extrapolation beyond the grid limits.

\subsection{\texorpdfstring{$\alpha_{S}$}{alpha-S} Function}

\begin{figure}[t] 
\centering{}
\centering{}\includegraphics[width=0.45\textwidth]{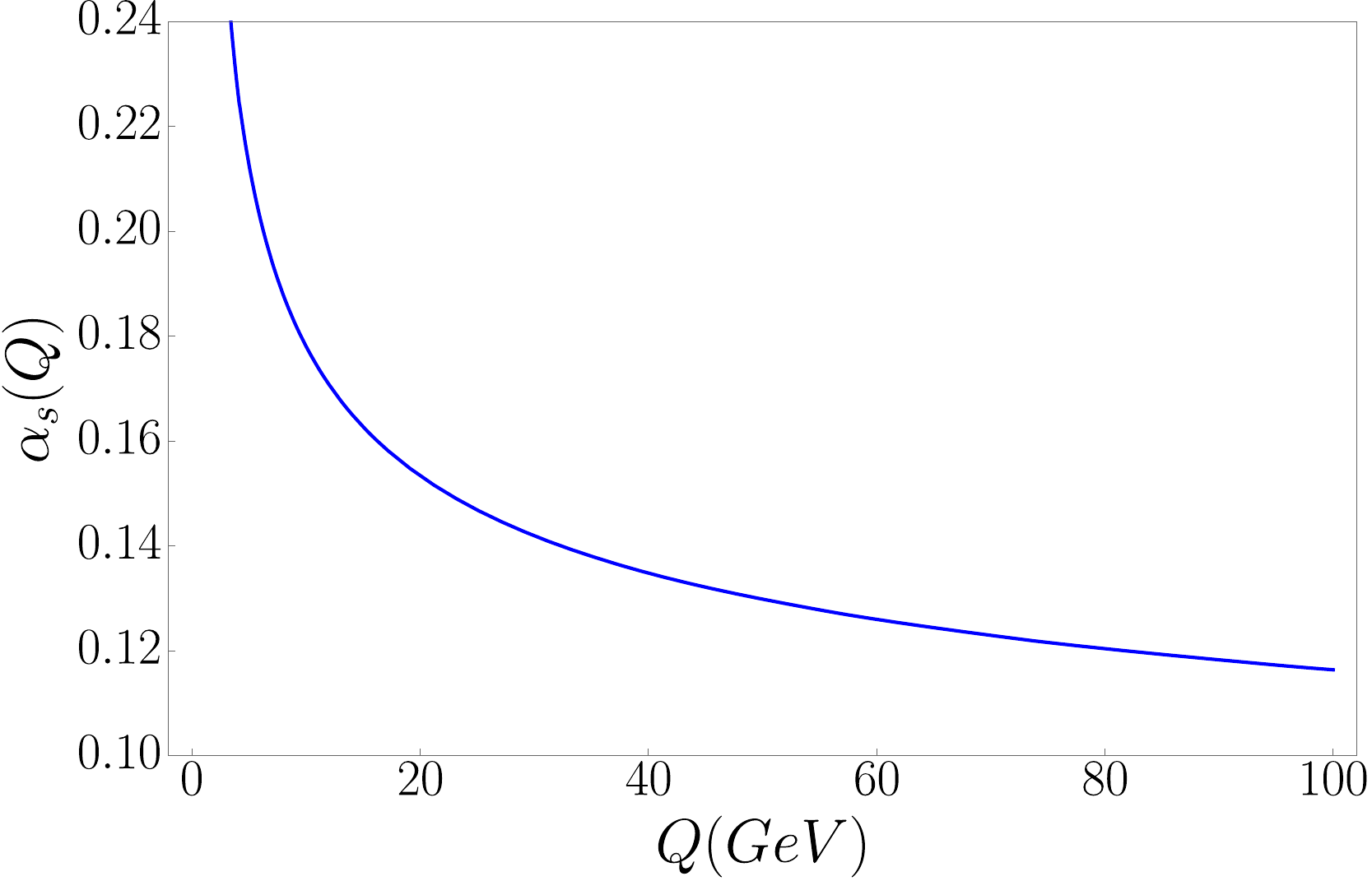}
\includegraphics[width=0.45\textwidth]{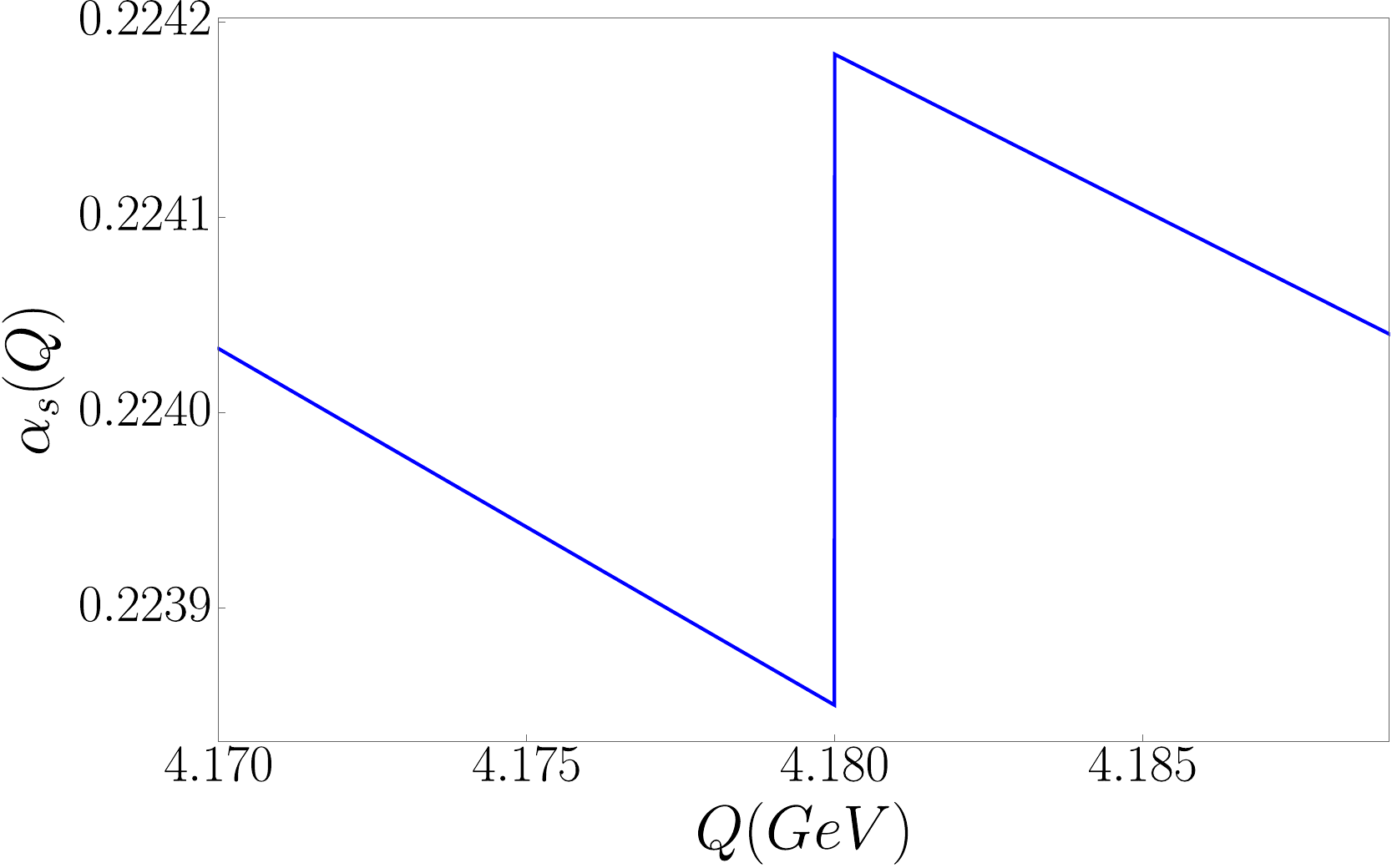}
\protect
\caption{$\alpha_{S}(Q)$ vs. $Q$ in GeV from \nnpdf. Note the discontinuity
across the $m_{b} = 4.18\ \mathrm{GeV}$ threshold which is enlarged in Fig.
b). 
\label{fig:alphas}}
\end{figure}

For some of the PDF sets, the value of $\alpha_{S}(Q)$ is provided
as a list of points associated with $Q_{vec}$. For these sets, we
interpolate
$\alpha_{S}(Q)$ to provide a 
matched function called \cmd{pdfAlphaS{[}iSet,Q{]}}; this is displayed in 
Fig.~\ref{fig:alphas} for a sample PDF 
set.\footnote{Since at Leading Order (LO), 
$\alpha_{S}(Q)=1/[\beta_{0}\ln(Q^{2}/\Lambda^{2})]$, we obtained improved 
results by interpolating in $1/\alpha_{S}(Q)$.} 
The \cmd{pdfGetInfo{[}iSet{]}} 
function will display the information associated with the corresponding PDF 
set (including any $\alpha_{S}$ values). If the PDF set does not have any 
$\alpha_{S}$ information, the \cmd{pdfAlpha} function will return 
\cmd{Null}. In Fig.~\ref{fig:alphas}-a) we display $\alpha_{S}(Q)$ 
for the \nnpdf\ set, and in Fig.~\ref{fig:alphas}-b) we enlarge the region
near $m_{b}=4.18\ \mathrm{GeV}$ to display the discontinuity. In general,
$\alpha_{S}(Q)$ will be discontinuous at NNLO and higher and at all mass
thresholds, $\{m_{c},m_{b},m_{t}\}$.

\section{Sample Plots \& Calculations\label{sec:Sample-Plots}}

The advantage of importing the PDF sets into Mathematica is that we
have the complete set of built-in tools that we can use for calculating
and graphing. We illustrate some of these features here.

\subsection{Graphical Examples}

\begin{figure}[t] 
\centering{}
\includegraphics[width=0.45\textwidth]{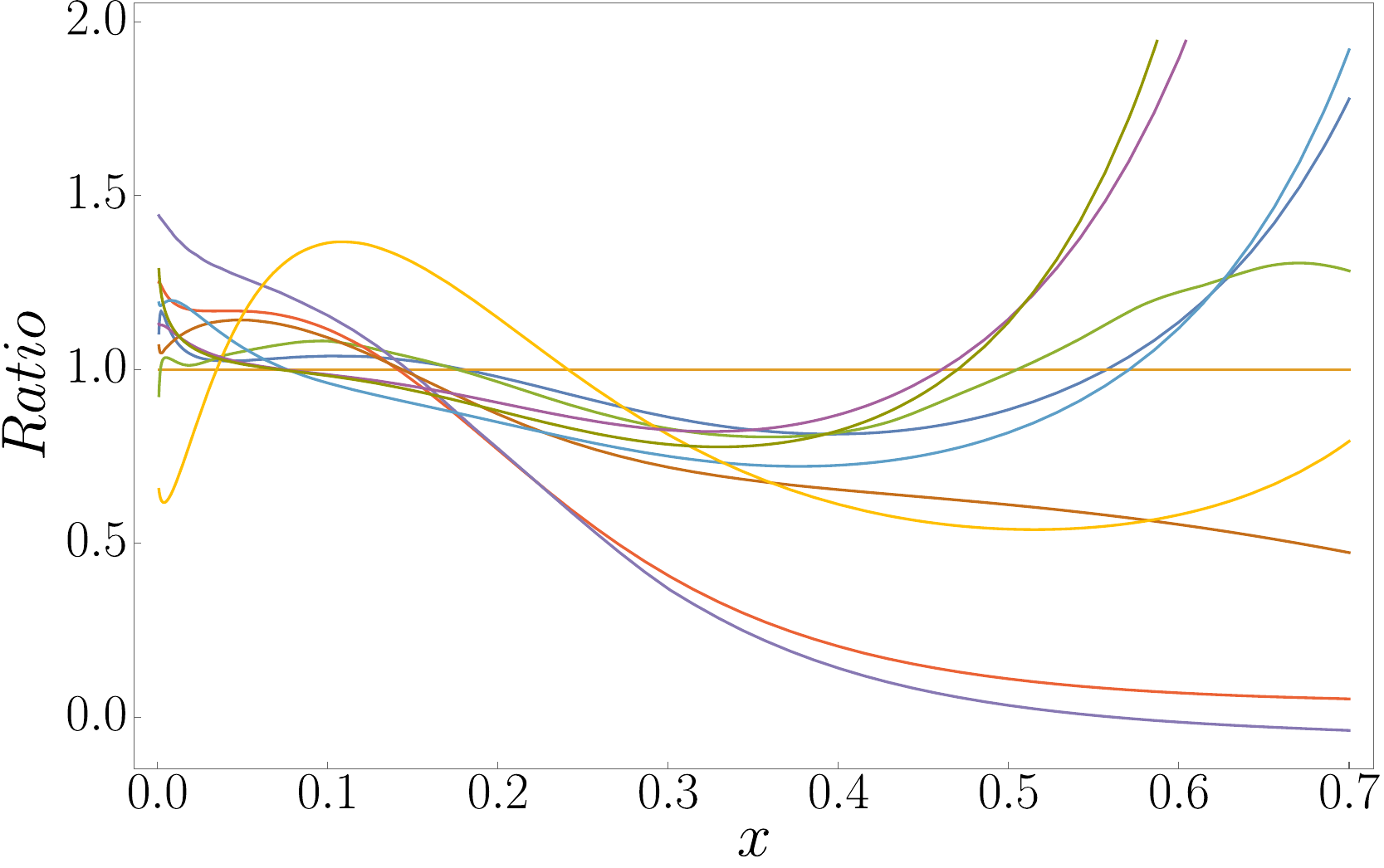}
\hfil
\includegraphics[width=0.45\textwidth]{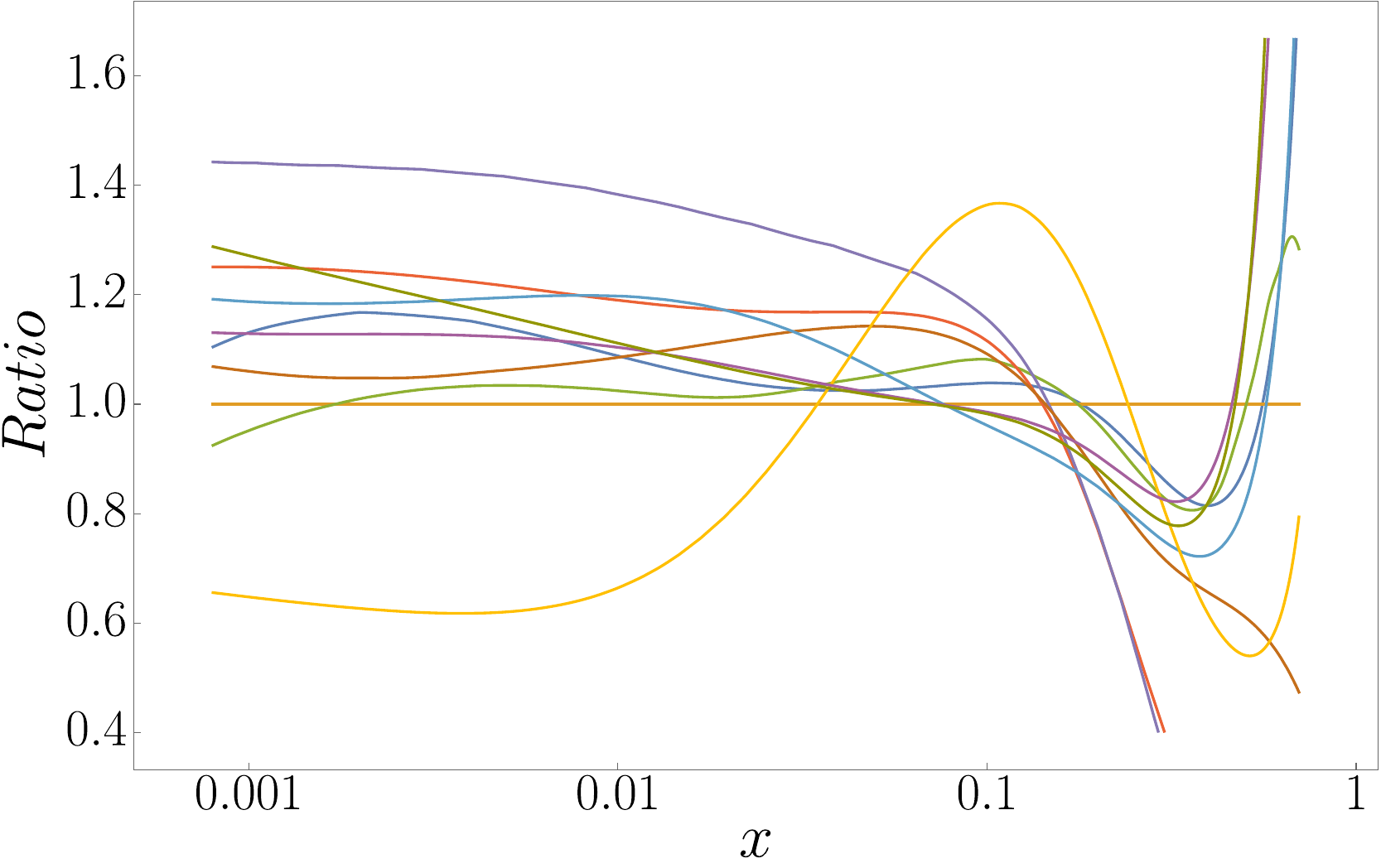}
\protect
\caption{Sample linear and log ratio plots of the gluon PDFs from Table~
\ref{tab:MomSum} compared to CT14 as a function of $x$ at 
$Q = 2.0\ \mathrm{GeV}$. 
\label{fig:LogLinear}}
\end{figure}

To highlight the graphical capabilities, in Figure~\ref{fig:LogLinear}
we display a selection of PDFs using both linear (left) and log
(right) scale. Using the flexible graphics capabilities of \mma\
it is easy to automatically generate such plots for different PDF
sets.

\subsection{Small {\it x} Extrapolation}

\begin{figure}[t] 
\centering{}
\includegraphics[width=0.45\textwidth]{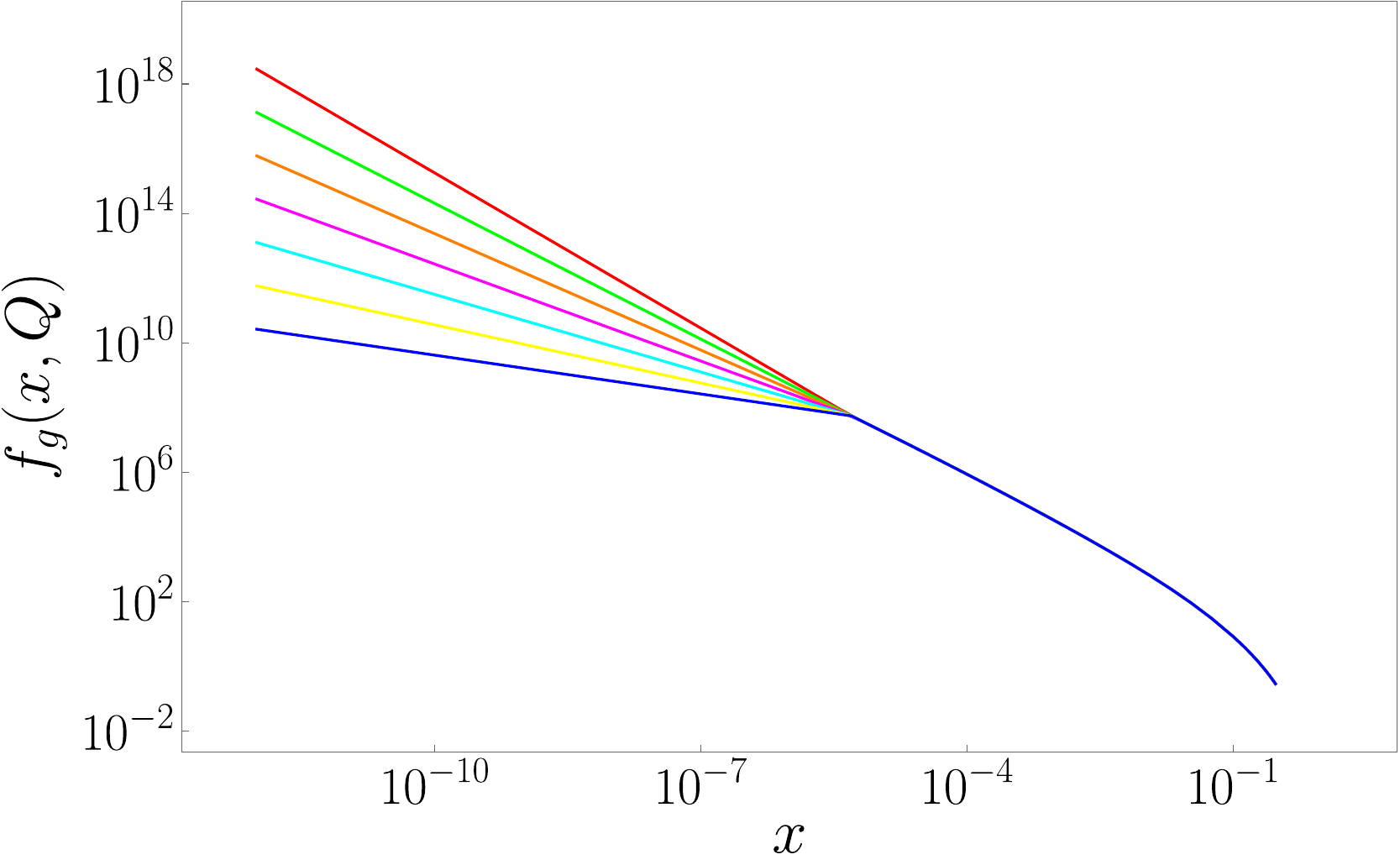}
\protect
\caption{Small $x$ extrapolation of the gluon PDF from the nCTEQ15 proton
at $Q = 100~\mathrm{GeV}$ using \cmd{pdfLowFunction}. Here, 
$x_{min}=5\times10^{-6}$, and the extrapolation exponent $1/x^{a}$ is set to 
$a=\{0.4,\,0.6,\,0.8,\,1.0,\,1.2,\,1.4,\,1.6\}$.
\label{fig:smallX}}
\end{figure}

Sometimes it is useful to extrapolate to low $x$ values beyond the
limits of the PDF grid; for example, the study of high energy cosmic
ray experiments that use very small $x$ extrapolations.\citet{Arguelles:2015wba,Bhattacharya:2015jpa} 
We provide the command 
\cmd{ pdfLowFunction{[}iSet,iParton,x,Q,power{]}} which allows the user to 
choose the extrapolation power in the small $x$ region.\footnote{The 
``\cmd{a}'' argument is optional; the default power is $1.0$. We use a 
separate function \cmd{pdfLowFunction} so as not to slow the computation of 
\cmd{pdfFunction}. } An example is displayed in Fig.~\ref{fig:smallX} 
for the nCTEQ15 proton PDF. The minimum $x$ value for this set for the grid is 
$x_{min}=5\times10^{-6}$; beyond this limit \cmd{pdfLowFunction} will 
extrapolate using the form $1/x^{a}$. In this example, we vary the power 
from $0.4$ to $1.6$; using the \mma\ integration routines it is easy to 
find that this range of variation in the small $x$ behavior will only
change the momentum fraction of the gluon by $1/2\%$. 


\begin{table*}[t] 
\renewcommand{\arraystretch}{1.5}
\centering{}
\begin{adjustbox}{max width=\textwidth}
\begin{tabular}{|c||c||c|c|c|c|c|c|c|c|c|}
\hline 
PDF Set  &~& Total  & $\overline{u}$  & $\overline{d}$  & $g$  & $d$  & $u$  
& $s$  & $c$  & $b$\tabularnewline
\hline 
\hline 
MSTW2008nnlo68cl \citep{Martin:2009iq}  &\crule[math1]& ~~99.87  & 3.3  & 3.8  & 43.5  & 14.6  
& 29.3  & 2.0  & 0.7  & 0\tabularnewline
\hline 
CT14nnlo \citep{Dulat:2015mca}  &\crule[math2]& 100.01  & 3.1  & 3.7  & 43.4  & 14.6  & 29.7  
& 2.0  & 0.8  & 0\tabularnewline
\hline 
{\small{}NNPDF30\_nnlo\_as\_0118\_nf\_6 \citep{Ball:2014uwa}}  &\crule[math3]& 99.98  & 3.2  
& 3.7  & 43.6  & 14.6  & 29.4  & 2.2  & 0.8  & 0\tabularnewline
\hline 
HERAPDF20\_NLO\_VAR \citep{Abramowicz:2015mha}  &\crule[math4]& 99.98  & 3.9  & 3.0  & 41.7  
& 14.6  & 31.2  & 2.2  & 0.6  & 0\tabularnewline
\hline 
abm12lhc\_5\_nnlo \citep{Alekhin:2013nda}  &\crule[math5]& 100.14  & 2.9  & 3.5  & 43.4  
& 14.8  & 30.4  & 2.0  & 0.7  & 0\tabularnewline
\hline 
CJ15nlo \citep{Accardi:2016qay}  &\crule[math6]& 99.96  & 3.0  & 3.7  & 43.3  & 15.1  
& 29.8  & 1.8  & 0.7  & 0\tabularnewline
\hline 
nCTEQ15\_1\_1 \citep{Kovarik:2015cma}  &\crule[math7]& 100.10  & 3.1  & 3.8  & 43.0  & 15.0  
& 30.2  & 1.8  & 0.7  & 0\tabularnewline
\hline 
nCTEQ15\_208\_82 \citep{Kovarik:2015cma}  &\crule[math8]& 99.99  & 2.7  & 3.4  & 44.6  
& 17.0  & 27.2  & 1.8  & 0.8  & 0\tabularnewline
\hline 
ct10.pds \citep{Lai:2010vv}  &\crule[math9]& 99.97  & 3.0  & 3.7  & 43.4  & 14.6  & 29.6  
& 2.2  & 0.7  & 0\tabularnewline
\hline 
ctq66m.pds \citep{Nadolsky:2008zw}  &\crule[math10]& 99.98  & 2.9  & 3.6  & 43.6  & 14.5  
& 29.4  & 2.3  & 0.7  & 0\tabularnewline
\hline 
\end{tabular}
\end{adjustbox}
\protect
\caption{We compute the momentum sum rule, Eq.~\ref{eq:sum}, (in percent)
for the individual partons at $Q = 3\ \mathrm{GeV}$. Partons $\{\bar{s},\bar{c},\bar{b}\}$
are not shown, but are equal to $\{s,c,b\}$. The totals sum to 100\%
within  uncertainties of integration and interpolation. Here the colors matched with each set correspond to 
that set in Fig.~\ref{fig:manyPDFs} and Fig.~\ref{fig:LogLinear}
\label{tab:MomSum} }
\end{table*}

\subsection{Momentum Sum Rules\label{sub:Momentum-Sum-Rules}}

\begin{figure}[t] 
\centering{}
\includegraphics[width=0.45\textwidth]{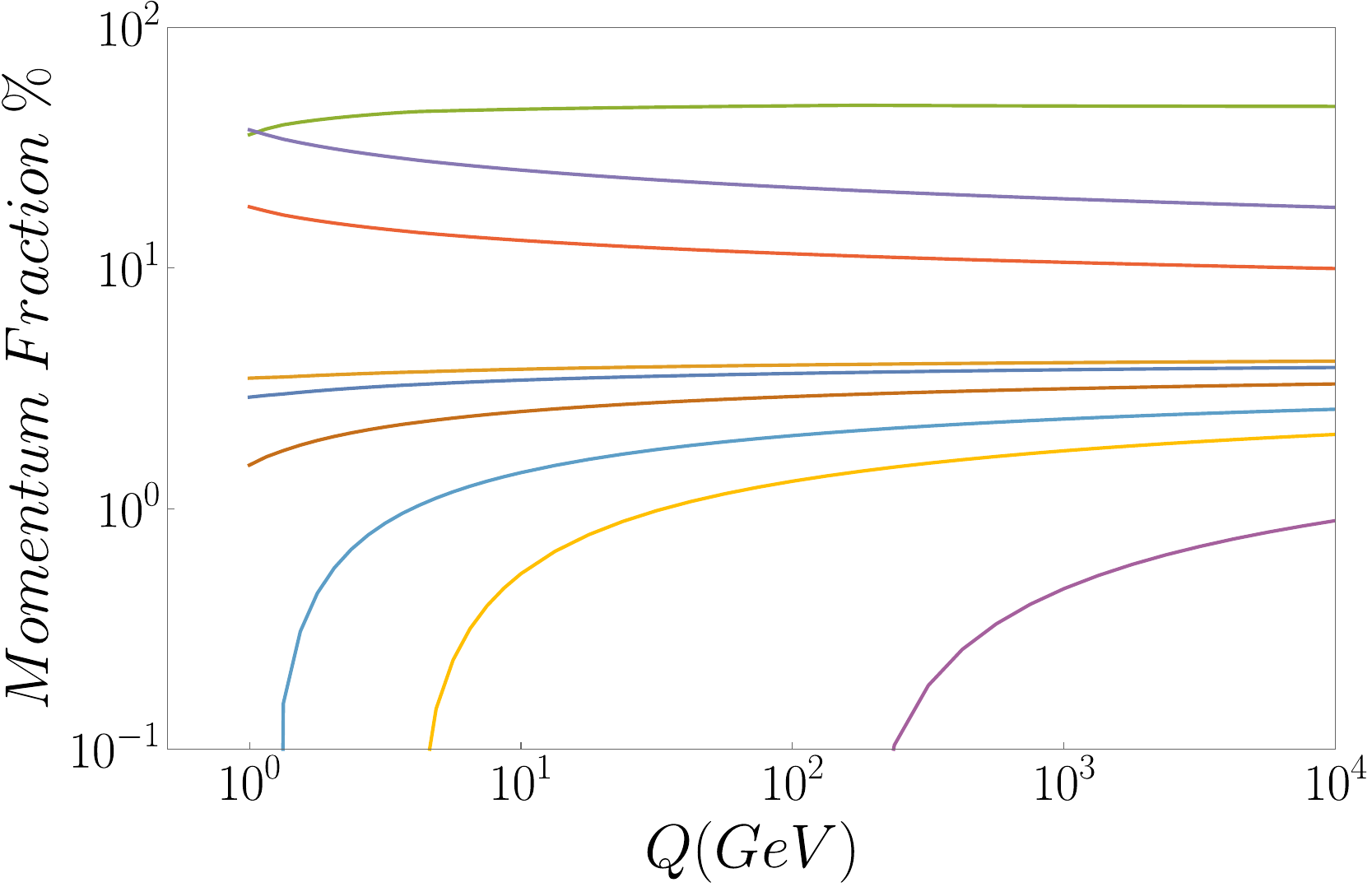}
\protect
\caption{The integrated momentum fraction Eq.~\ref{eq:sum} of the
PDF flavors vs.\  $Q$ in GeV for the NNPDF set. At large $Q$ the curves
are (in descending order) $\{g, u,d, \bar{u}, \bar{d}, s,c, b, t\}$.
\label{fig:IntMom}}
\end{figure}

The PDFs satisfy a number of momentum and number sum rules, and this
provides a useful cross check on the results. The momentum sum rule:
\begin{equation}
\sum_{i}\int_{0}^{1}dx\,x\,f_{i}(x,Q)=1,
\label{eq:sum}\\
\end{equation}
says that the total momentum fraction of the partons must sum to 100\%. If any single parton flavor
were not imported correctly, this cross-check would be violated; hence,
this provides a powerful ``sanity check'' on our implementation. In
Table~\ref{tab:MomSum} we display the partonic momentum fractions
(in percent) and the total; for each PDF set the momentum sum rule
checks within numerical accuracy.\footnote{Numerical uncertainties arise from 
the extrapolation down to $x\to0$, the interpolation, and the integration 
precision.} 

While Table~\ref{tab:MomSum} presented the momentum fraction for
a single $Q$ value ($3\ \mathrm{GeV}$), it is interesting to see how these values
change with the energy scale. In Fig.~\ref{fig:IntMom} we show the
momentum carried by each PDF flavor (in percent) as a function of
$Q$ in GeV. We can see the heavy quarks, $\{c,b,t\}$ enter as we
cross the flavor mass thresholds. In the limit of large $Q$, the $\{\bar{u},\bar{d},
\bar{s}\}$ PDFs approach each other asymptotically.

\subsection{Nuclear Correction Factors}

\begin{figure}[t] 
\centering{}
\includegraphics[width=0.45\textwidth]{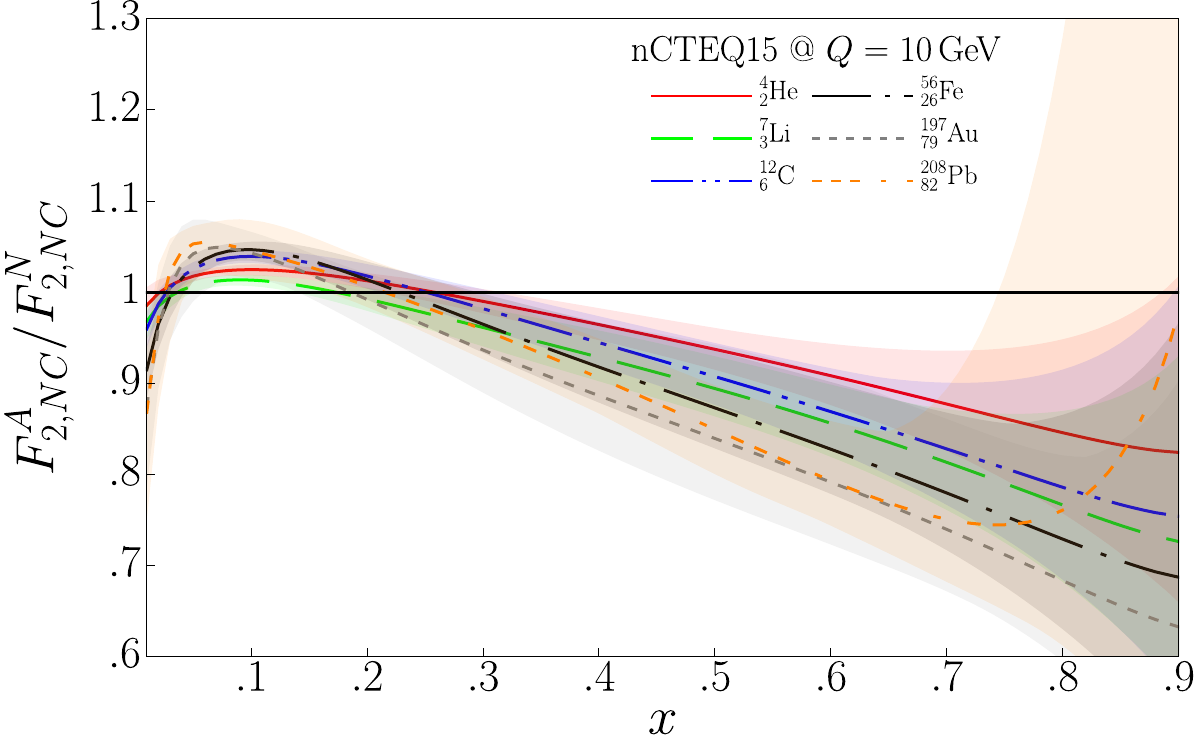}
\hfil
\includegraphics[width=0.45\textwidth]{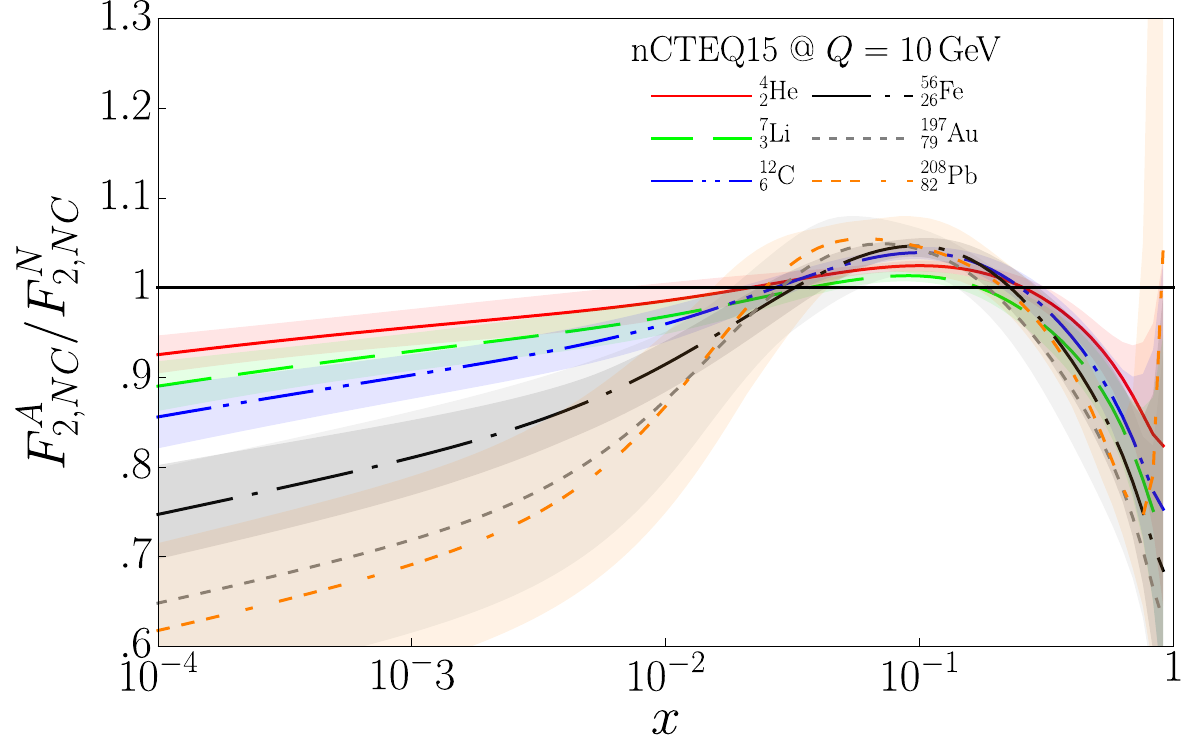}
\protect
\caption{Nuclear correction ratios $F_{2}^{A}/F_{2}^{N}$ vs. $x$ for 
$Q=10~\mathrm{GeV}$ for the nCTEQ15 PDFs over an iso-scalar target.
The left plot is on a linear scale, 
and the right plot is a log scale. This figure is comparable to Fig.~1 of 
Ref.~\cite{Schienbein:2009kk}.
\label{fig:NucRatios}
}
\end{figure}

Given the PDFs, it is then trivial to build up simple calculations.
In Fig.~\ref{fig:NucRatios} we display the nuclear correction factors
$F_{2}^{A}/F_{2}^{N}$ for a variety of nuclei. Here, the $F_{2}$
structure functions are related to the PDFs via $F_{2}^{A}(x,Q)=x\sum_{q}\,
e_{q}^{2}\,f_{q/A}(x,Q)$ at leading order where $F_{2}^{N}$ is an isoscalar, and $F_{2}^{A}$ is 
the scaled structure function\footnote{More specifically, $F_{2}^{N}$ is the 
average of the proton and neutron $(p+n)/2$ and $F_{2}^{A}$ is composed of Z 
protons, (A-Z) neutrons, and scaled by A to a make it ``per nucleon:'' 
$\left[Z\,p+(A-Z)\,n\right]/A$.} for nuclei $A$. We have also superimposed the 
uncertainty bands; we will discuss this more in the following Section.

\subsection{Luminosity}

\begin{figure}[t]  
\begin{centering}
\includegraphics[width=0.45\textwidth]{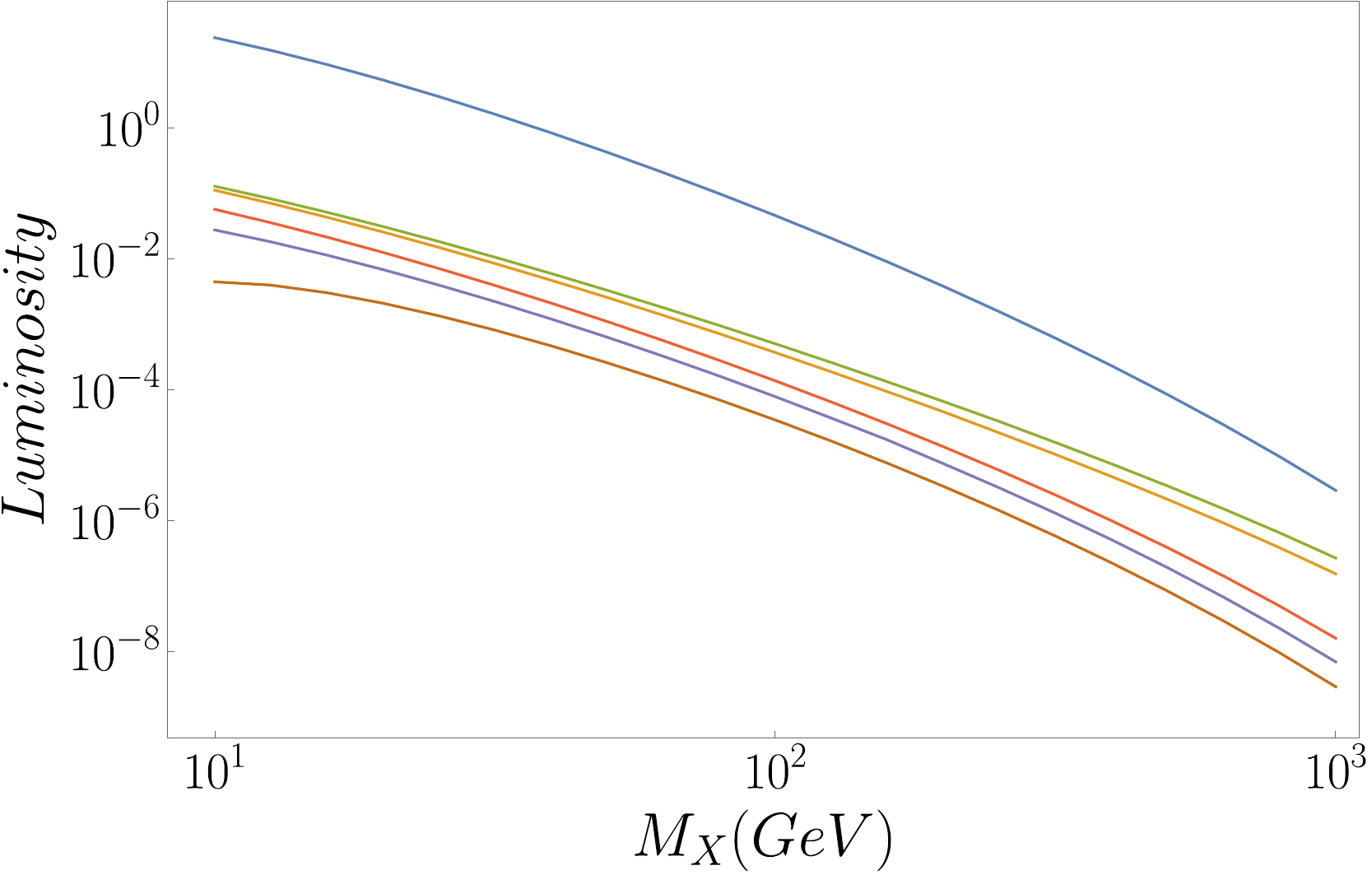}
\par
\end{centering}
\protect
\caption{The differential parton-parton luminosity $ d L_{a \bar{a}}/d M_{X}^{2} $ vs.
$ M_{x} $ in GeV at $ \sqrt{s} = 14\,\mathrm{TeV}$ for (in descending order) 
$a = \{g,u,d,s,c,b\} $.
\label{fig:lumi}}
\end{figure}

Using the integration capabilities of \mma\ it is easy to compute
the differential parton-parton 
luminosity\footnote{There are other definitions of the 
luminosity in the literature which are dimensionless such as ${\cal L}=f_a \otimes f_b$.}
for  partons $a$ and $b$:\citet{Campbell:2006wx} 
\begin{equation}
\frac{d {\cal L}_{ab}}{d \hat{s}}
	= \frac{1}{s}\frac{1}{1+
	\delta_{ab}}\int_{\tau}^{1}\frac{dx}{x}\,f_{a}(x, \sqrt{\hat{s}})\,f_{b}(
 	\frac{\tau}{x}, \sqrt{\hat{s}})\,+(a\leftrightarrow b),
\label{eq:lum}\\
\end{equation}
where $\tau = \hat{s} / s$, and the cross section is 
\begin{equation}
\sigma = \sum_{a,b} \int \left( \frac{d \hat{s}}{\hat{s}} \right)
	\left( \frac{d {\cal L}_{ab}}{d \hat{s}} \right)
	(\hat{s}\  \hat{\sigma}_{ab}).
\label{eq:lum2}\\
\end{equation}
Note, the  luminosity  definition  of  Eq.~\ref{eq:lum} 
has   dimensions of a cross section ($1/\hat{s}$),
and in Eq.~\ref{eq:lum2} we multiply by a scaled (dimensionless) cross section 
($\hat{s} \, \hat{\sigma}_{ab}$).

We define the \cmd{pdfLuminosity} function
to compute Eq.~\ref{eq:lum}. The hadron-hadron production cross section for producing
 particle of mass $\sqrt{\hat{s}} = M_{X}$ is proportional to the luminosity times
the scaled partonic cross section $\hat{s} \sigma$ as in Eq.~\ref{eq:lum2}. 
In Fig.~\ref{fig:lumi} we display the
differential luminosity $ d L_{a\bar{a}}/ d M_{X}^{2}$ for parton--
anti-parton ($a\bar{a}$) combinations; this luminosity would be appropriate if
we were interested in estimating the size of the cross section for 
the process of quark--anti-quark annihilation
into a Higgs boson, $b\bar{b}\to H$ for example.\footnote{\Mane\ also has the capability 
to handle custom PDFs. This allows the user to explore a wide variety of phenomena, 
such as intrinsic heavy quarks, as long as the custom PDFs are written in either \lha\ or \cteq\ format}

\subsection{W Boson Production}

\begin{figure}[t] 
\centering{}
\includegraphics[width=0.45\textwidth]{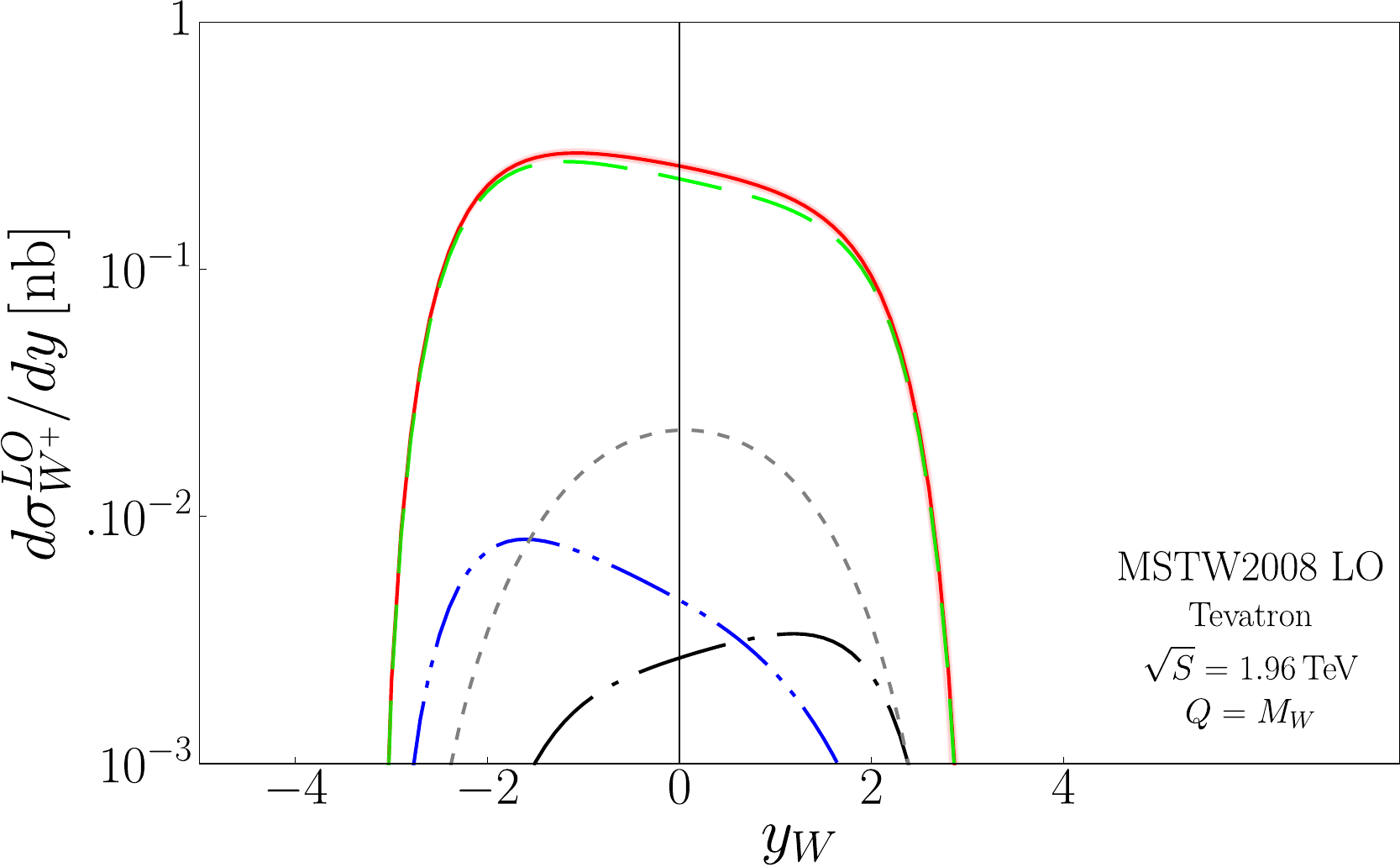}
\hfil
\includegraphics[width=0.45\textwidth]{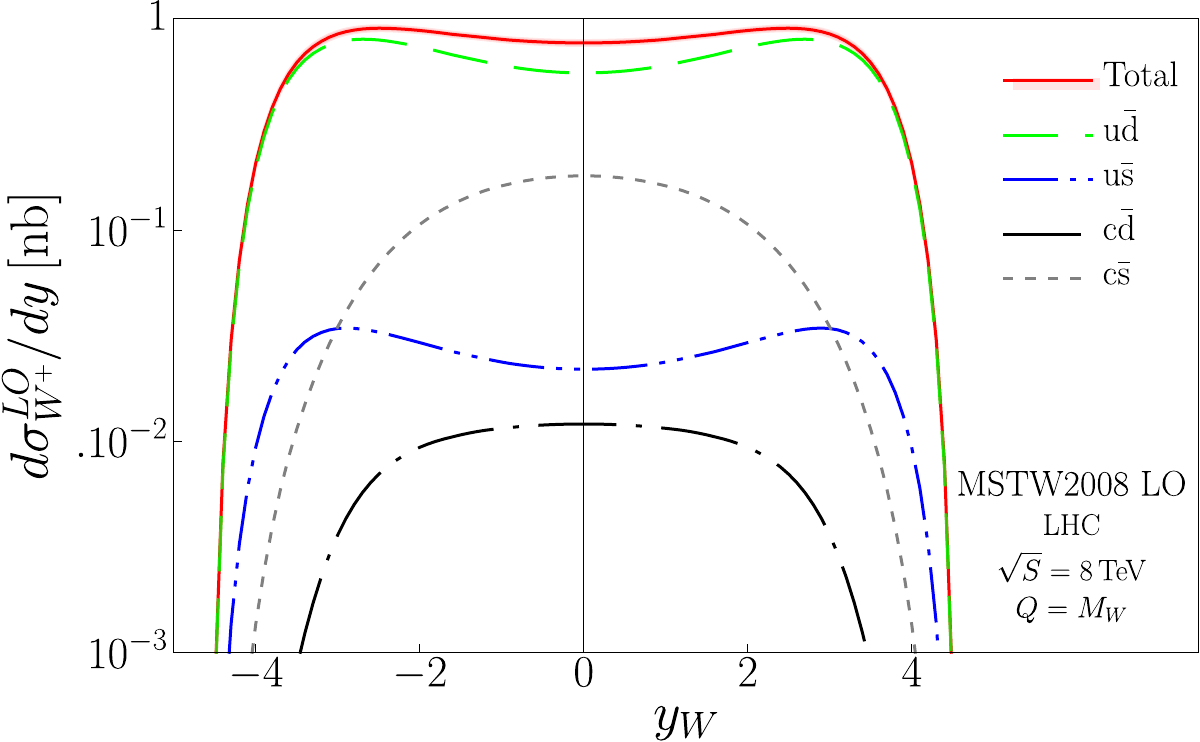}
\protect
\caption{Leading-Order $W^{+}$ production cross section, $d\sigma/dy$ at the 
Tevatron ($p\bar{p},$~$ 1.96\ \mathrm{TeV} $) and the LHC ($pp$,~$ 8\
\mathrm{TeV}$). We display the total cross section and the individual partonic 
contributions. 
\label{fig:wprod}}
\end{figure}

Next, we compute a simple leading-order (LO) cross section for $W^{+}$ boson 
production at the Tevatron proton--anti-proton collider ($ 1.96\
\mathrm{TeV} $) and the LHC proton-proton collider ($ 8\ \mathrm{TeV} $). 
Schematically, the cross section is $\sigma(W^{+})=f_{a}\otimes f_{b}
\otimes\omega_{ab\to W^{+}}$. There are two convolution integrals, but the 
constraint that the partonic energies sum to the boson mass $W^{+}$ eliminates 
one.\citep{Kusina:2012vh,Kovarik:2012te} Hence, this can easily be performed inside of  
\mma, and the results are displayed in Fig.~\ref{fig:wprod}. It is 
interesting to note the much larger width of the rapidity distribution at the
LHC as well as the increased relative contribution of the heavier
quark channels (such as $c\bar{s}$ and $u\bar{s}$).

\section{Error PDFs \& Correlations\label{sec:Error-PDFs:}}

\subsection{PDF Uncertainties}

\begin{figure}[t]  
\centering{}
\includegraphics[width=0.45\textwidth]{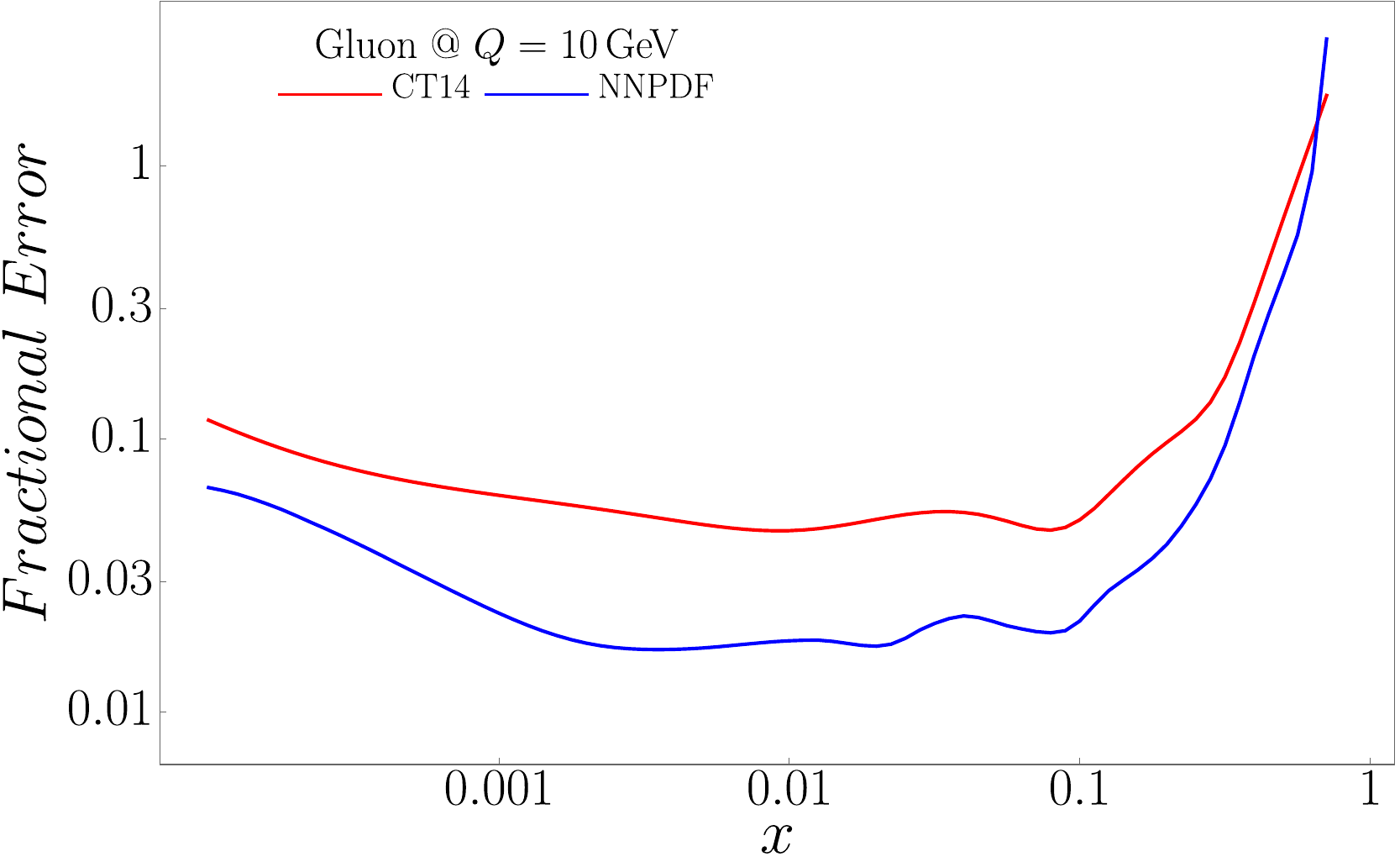}
\hfil
\includegraphics[width=0.45\textwidth]{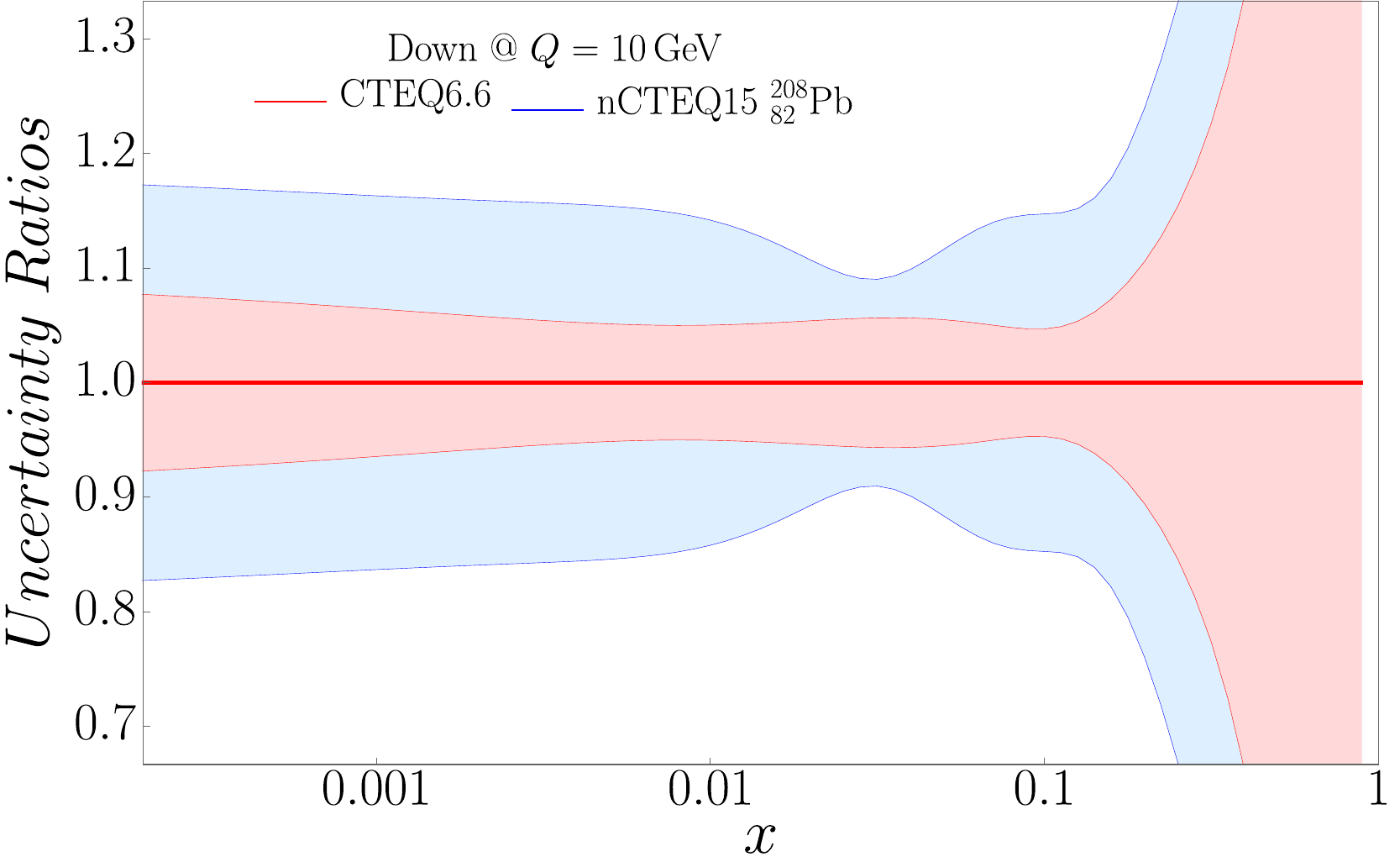}
\protect
\caption{The fractional PDF uncertainty vs. $x$ at $Q=10\ \mathrm{GeV}$. a) The
upper (red) curve is CT14 using the \cmd{pdfHessianError}
function, and the lower (blue) curve is the \nnpdf\ using the \cmd{pdfMCError}
function for the gluon. (Note, these curves do not necessarily represent the same
confidence level.) b) The down quark PDF uncertainty band for the CTEQ6.6
PDFs (inner, red) and the \cmd{nCTEQ15} lead 208 (outer, blue); 
\label{fig:errorPDF}}
\end{figure}

We now examine some of the added features provided by the \cmd{pdfError}
module. 
To accommodate the PDF errors, it is common for the PDF groups to
release a set of grids to characterize the uncertainties; the number
of PDFs in each error is typically in the range 40 to 100, but can
in principle be as many as 1000. 

As \mma\ handles lists naturally, we can exploit this feature
to manipulate the error PDFs. The \cmd{pdfFamilyParseLHA}
and \cmd{pdfFamilyParseCTEQ} functions will read an entire
directory of PDFs and return the associated set numbers as a list;
this list can then be used to manipulate the entire group of error
PDFs. 

For example, we can use this feature to read the 100 PDFs of the \nnpdf\
set displayed in Fig.~\ref{fig:errorPDF}, capture the returned list
of \cmd{iSet} values, and pass this to the plotting function; we'll
describe this more in the following.

When working with the error PDFs, the first step is to take the list
of \cmd{iSet} values and obtain a list of the PDF values. 
Constructing the PDF error depends on whether the set is based on
the Hessian or the Monte Carlo method. 

The Hessian PDF error sets can be organized as follows 
$\{X_{0}$,\, 
$X_{1}^{+}$,\, 
$X_{1}^{-}$,\, 
$X_{2}^{+}$,\, 
$X_{2}^{-}$, 
$...$,\, 
$X_{N}^{+}$,\, 
$X_{N}^{-}\}$ where $X_{0}$ 
represents the central set, $\{X_{1}^{+},\,X_{1}^{-}\}$ represent the plus and 
minus directions along eigenvector \#1, and so on up to eigenvector $N$. For 
the Hessian PDF sets, there should be an odd number equal to $2N+1$ where N is 
the number of eigenvector directions. The PDF errors can then be constructed 
using symmetric, plus, or minus definitions:\citep{Pumplin:2002vw,Campbell:2006wx}
\begin{eqnarray}
\Delta X^{\mathit{Hess}}_{\mathit{sym}} & = & \frac{1}{2}\sqrt{\sum_{i=1}^{N}\left[X_{i}^{+}-X_{i}^{-}
\right]^{2}}\label{eq:sym}\\
\Delta X^{\mathit{Hess}}_{\mathit{plus}} & = & \sqrt{\sum_{i=1}^{N}\left[\mbox{\ensuremath{\max}}\left
\{ X_{i}^{+}-X_{0},\,X_{i}^{-}-X_{0},\,0\right\} \right]^{2}}\label{eq:plus}\\
\Delta X^{\mathit{Hess}}_{\mathit{minus}} & = & \sqrt{\sum_{i=1}^{N}\left[\mbox{\ensuremath{\max}}
\left\{ X_{0}-X_{i}^{+},\,X_{0}-X_{i}^{-},\,0\right\} \right]^{2}}\label{eq:minus}
\end{eqnarray}
These can be computed using the function 
\cmd{pdf\-Hessian\-Error\-[iSet,(method)]}, 
and can take an optional ``\cmd{method}'' argument, 
\{\cmd{``sym'',}\cmd{``plus'',}\cmd{``minus''}\}, to specify which formula is used to 
compute the error; the default being \cmd{``sym''}. 

We next turn to the Monte Carlo sets. For example, the \nnpdf\ set (\#3
in Table~\ref{tab:MomSum}) has 101 elements; the ``zeroth'' set is the
central set, and the remaining 100 replica sets span the PDF uncertainty space.
The central set is the average of all the sets, and the PDF error
is the standard deviation of the 100 replica sets. For these sets, 
\cmd{pdfMC\-Central} will return the central PDF value.
\cmd{pdfMC\-Error\-[iSet,(method)]} will return the associated error. This 
function can also take an optional ``\cmd{method}'' argument, \{\cmd{``sym'',
``plus'',``minus''}\}, defined by 
Eqs.~\ref{eq:MCsym}, \ref{eq:MCplus} and~\ref{eq:MCminus}.\citet{Alekhin:2011sk,Gao:2013bia} 

The modification from the Hessian case is due to the MC error PDFs using 
replica sets, not eigenvector pairs.\footnote{See the \lha\ 
reference\citet{Buckley:2014ana} for a more complete description of the error 
definitions and calculation.}
The formula for $\Delta X^{MC}_{sym}$ is a 
straightforward extension of the Hessian case: 
\begin{eqnarray}
\Delta X^{\mathit{MC}}_{\mathit{sym}} & = & \sqrt{ \frac{1}{N_{rep}}\sum_{i=1}^{N}
\left[X_{i}-X_{0}
\right]^{2}} \quad . 
\label{eq:MCsym}
\end{eqnarray}
where $N_{rep}$ counts the 100 replica sets not including the ``zeroth'' central set.
This quantity is simply the standard deviation of the values. 
The  $1/\sqrt{N_{rep}}$ factor compensates for the fact that Monte Carlo 
sets can have an arbitrary number of replicas, in contrast to the Hessian
sets which have a fixed number of eigenvector sets. 

It is possible to define extensions for Monte Carlo 
``plus'' and ``minus'' uncertainties as:\cite{Nadolsky:2001yg}
\begin{eqnarray}
\Delta X^{\mathit{MC}}_{\mathit{plus}} & = & \sqrt{\frac{1}{N^{+}_{rep}}\sum_{i=1}^{N}
\left[\mbox{\ensuremath{\max}}
\left
\{ X_{i}-X_{0},\,0
\right\} 
\right]^{2}}\label{eq:MCplus}
\\
\Delta X^{\mathit{MC}}_{\mathit{minus}} & = & \sqrt{\frac{1}{N^{-}_{rep}}\sum_{i=1}^{N}
\left[\mbox{\ensuremath{\max}}
\left\{ X_{0} -  X_{i}   ,\,0
\right\} 
\right]^{2}} \quad ,
\label{eq:MCminus}
\end{eqnarray}
where $N^{\pm}_{rep}$ are the number of replicas above/below the mean. 

In Fig.~\ref{fig:errorPDF}-a), we compute the fractional PDF error
for the CT14 PDF gluon using the \cmd{pdfHessianError} function
with the \cmd{``sym''} formula of  Eq.~\ref{eq:sym}. The same is done for the
\nnpdf\ set \cmd{pdfMCError} function, using Eq.~\ref{eq:MCsym}. As expected, we see
the uncertainty increase both as $x \to 1$ and at very small $x$ values. 

In Fig.~\ref{fig:errorPDF}-b), we compute the error bands for the down quark 
in the CTEQ6.6 proton PDF and also the \cmd{nCTEQ15} lead-208 PDF; as expected, we
see the uncertainties on the nuclear PDF are larger than the proton PDF 
uncertainties.

\subsection{Correlation Angle}

\begin{figure*}[t]  
\centering{}
\includegraphics[width=0.45\textwidth]{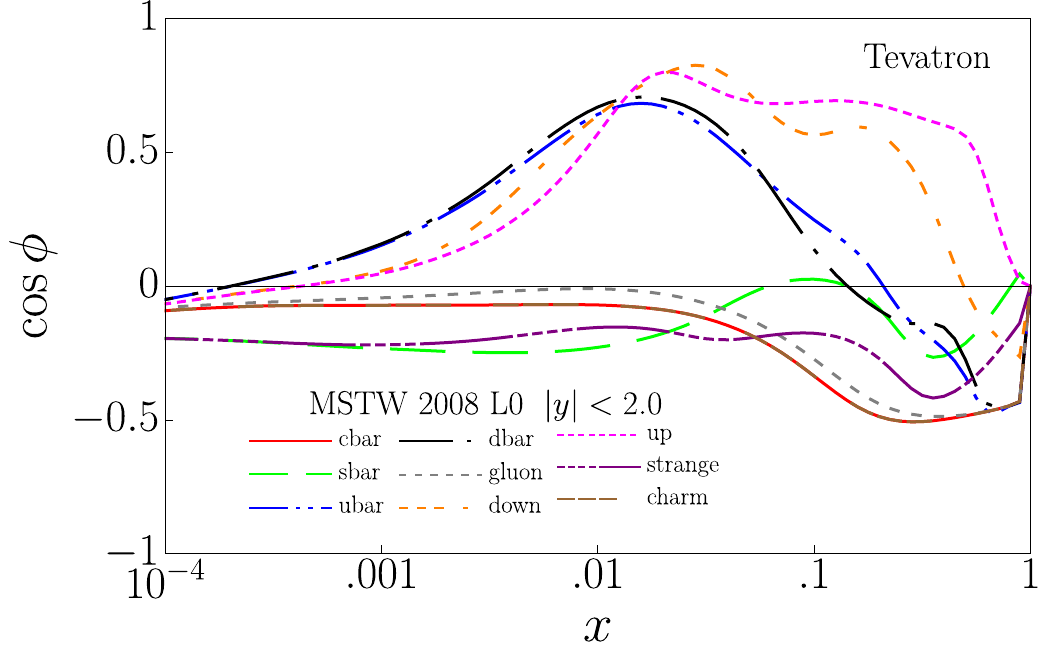}\hfill
\includegraphics[width=0.45\textwidth]{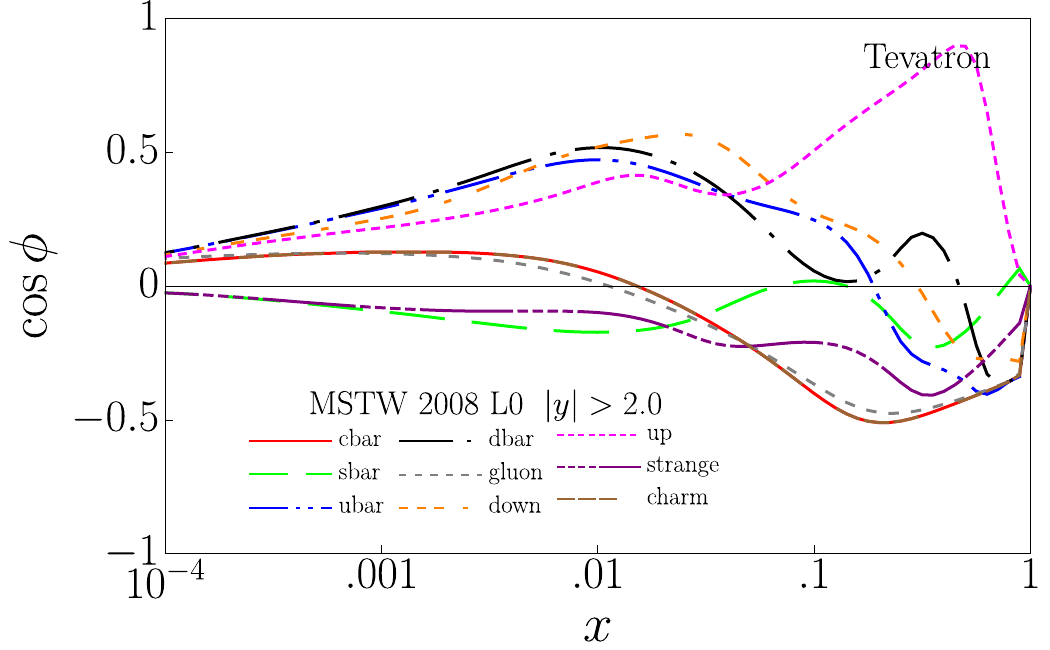}
\includegraphics[width=0.45\textwidth]{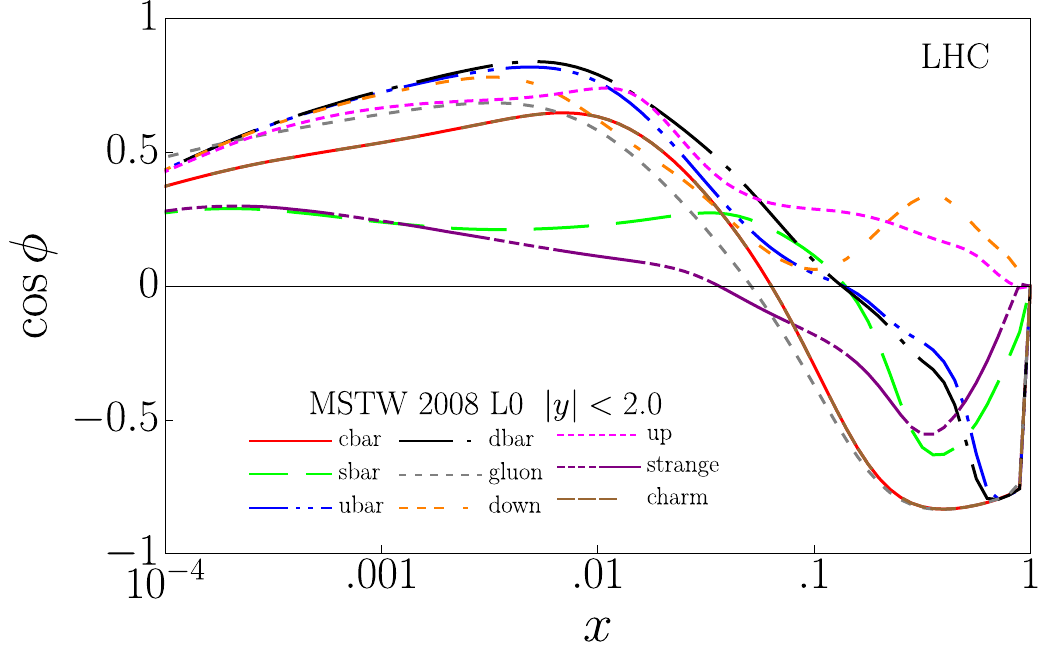} \hfill
\includegraphics[width=0.45\textwidth]{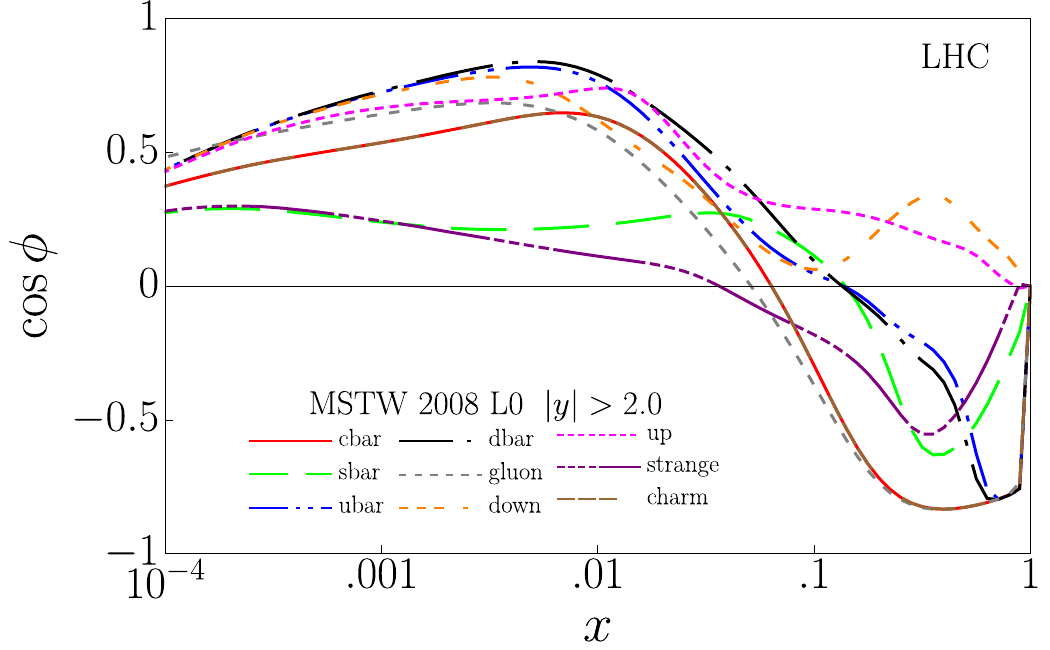}
\protect\caption{The cosine of the correlation angle, $\cos{\phi}$, as in 
Eq.~\ref{eq:cosphi}, as a function of Bjorken-$x$ for the leading-order $W^+$ cross 
section and each of the partonic flavors. Differences between Tevatron (top)
$p \bar{p}$ collisions at $\sqrt{S} = 1.96\ \mathrm{TeV}$ and LHC (bottom) $pp$
collisions at $\sqrt{S} = 8 \mathrm{TeV}$ are visible in the high-$x$ 
region in both the central ($\lvert y \rvert < 2 $) and high absolute 
rapidity ($ \lvert y \rvert > 2 $) regions.
\label{fig:correlation}}
\end{figure*}

Finally, we can compute the correlation cosines via the relation:
\citep{Nadolsky:2008zw} 
\begin{eqnarray}
\cos\varphi&=&
\frac{\overrightarrow{\nabla}X\cdot\overrightarrow{\nabla}Y}{\triangle X\,\triangle Y}
\nonumber \\
&=&\frac{1}{4\triangle X\,\triangle Y}\;
\sum_{i=1}^{N}
\left(X_{i}^{+}-X_{i}^{-}\right)\left(Y_{i}^{+}-Y_{i}^{-}\right)\;.
\label{eq:cosphi}
\end{eqnarray}
We have implemented separate functions \cmd{pdfHessianCorrelation}
and \linebreak \cmd{pdfMCCorrelation} as the computation of the uncertainty
in the denominator $\Delta X\,\Delta Y$ could depend on 
Eqs.~\ref{eq:sym}, \ref{eq:plus} and~\ref{eq:minus} 
or Eqs.~\ref{eq:MCsym}, \ref{eq:MCplus} and~\ref{eq:MCminus}.

In Fig.~\ref{fig:correlation} we display an example where we show the 
correlation cosine between the $W^+$ cross section and the partonic flavors 
for both the Tevatron and LHC. 
We observe the behavior of the flavors is quite similar except 
for the $u$ and $d$ quarks which stand out at large~$x$.

The cosine of the correlation angle indicates the degree to 
which the error on a particular 
parton's PDF contributes to the uncertainty on some function of the PDFs, usually a 
physical observable. A value close 
to one for some parton indicates that the PDF error on the observable is being 
driven by the error on that parton's PDF. Similarly, a value close to zero 
indicates that the error on the parton's PDF does not contribute significantly 
to the error on the observable. More details can be found 
in Ref.~\cite{Nadolsky:2008zw}.

\section{Conclusions\label{sec:Conclusions}}

We have presented the \Mane\ package which provides PDFs within the
\mma\ framework.  This is designed to work with any of the \lha\ format PDFs,
and is extensible to other formats such as the CTEQ PDS format.
\Mane\ can also work with nuclear PDFs such as the nCTEQ15 sets.

The \Mane\ package implements a number of novel features.  It adapts
YAML relations into \mma\ rules including unknown keys, and can
handle discontinuities in both the PDFs and $\alpha_s(Q^2)$.  We have
implemented a flexible interpolation with a tunable parameter, and it
can extrapolate to small $x$ with a variable power.
Additionally, we have implemented functions to facilitate the calculation
of PDF uncertainties for both Hessian and Monte Carlo PDF sets.

\Mane\ provides many tools to simplify calculations involving PDFs, and
is fast enough such that even one or two convolutions can easily be
computed within the \mma\ framework.  We illustrated these features with
examples of $W$ production, luminosity calculations, nuclear
correction factors, and $N_F$-dependent PDFs.

In summary, the \Mane\ package is a versatile, flexible,
user-extensible tool that can be used by beginning users to make
simple PDF plots, as well as by advanced users
investigating subtle features of higher-order discontinuities and PDF
uncertainty calculations.

\section*{Acknowledgments\label{sec:Acknowledgments}}

The authors would like to thank
V.~Bertone,
M.~Botje, 
A.~Buckley, 
P.~Nadolsky, 
and
V.~Radescu 
for help and suggestions. 
We are also grateful to our nCTEQ colleagues 
K.~Kovarik, 
A.~Kusina, 
T.~Jezo, 
C.~Keppel, 
F.~Lyonnet, 
J.G.~Morfin, 
J.F.~Owens, 
I.~Schienbein, 
\& 
J.Y.~Yu
for valuable discussions, 
testing the program, and providing useful feedback.

We acknowledge the hospitality of CERN, DESY, Fermilab, and KITP where a
portion of this work was performed. 
We thank HepForge for hosting the project files. 
This work was also partially supported 
by the U.S. Department of Energy under Grant No.  DE-SC0010129
and  by the National Science Foundation under Grant No. NSF PHY11-25915.

\newpage
\appendix

\addcontentsline{toc}{section}{Appendices}
\addtocontents{toc}{\protect\setcounter{tocdepth}{-1}}
\section{Appendices}

\subsection{\Mane\ Distribution Files\label{sec:ManeParse-Distribution-Files}}

\addtocontents{toc}{\protect\setcounter{tocdepth}{3}}
\addcontentsline{toc}{subsection}{\Mane\ Distribution Files\label{sec:ManeParse-Distribution-Files}}
\addtocontents{toc}{\protect\setcounter{tocdepth}{-1}}

The \Mane\ package is distributed as a gzipped tar file (about 2.6Mb),
and this is available at \cmd{cteq.org} or  \cmd{ncteq.HepForge.org}.

When this is unpacked, the \cmd{ManeParse} modules
\{\cmd{pdfCalc}, 
\cmd{pdfErrors},  
\cmd{pdfParseCTEQ}, 
\cmd{pdfParseLHA}\}  
will be in the \cmd{./MP\_Packages/} \hbox{directory.}


There is a \cmd{Demo.nb} \mma\ notebook which will illustrate
the basic functionality of the program; we also include
a  \cmd{Demo.pdf} file so the user can see examples of the correct output. 

We do not distribute any PDF
files, so these must be obtained from the \lha\ 
website\footnote{
\href{http://lhapdf.hepforge.org/}{http://lhapdf.hepforge.org/}}
or the CTEQ website.\footnote{
\href{http://cteq.org/}{http://cteq.org/}}\ 
The \cmd{READ\-ME} file will explain how to run the \cmd{MakeDemo.py}
python script to download and set up the necessary directories for
the PDF files.\footnote{Python is not essential to \Mane\ as the files can 
be setup manually.} 

The \cmd{MakeDemo.py} script will also run the \perl\ script \cmd{noe2.perl}
on the CT10 data files. Older versions of these files use a two digit
exponent (e.g. \mbox{1.23456E-12}), but occasionally three digits
are required in which case the value is written as \mbox{1.23456-123}
instead of \mbox{1.23456E-123}. While the GNU compiler writes and
reads this properly, other programs (including \mma) do not,
so the \cmd{noe2.perl} script fixes this. This script can also
be run interactively, in which case it will print out any lines that
are modified. 

There is a manual in both \mma\ format (manual\_v1.nb)
and PDF format (manual\_v1.pdf); this allows the user to execute the
notebook directly, but also see how the output should look. The manual
provides examples of all the functions of \Mane. 

There is also a glossary file
\cmd{User.pdf} which provides a list and usage of all the commands.


\subsection{A Simple Example}
\addtocontents{toc}{\protect\setcounter{tocdepth}{3}}
\addcontentsline{toc}{subsection}{A Simple Example}
\addtocontents{toc}{\protect\setcounter{tocdepth}{-1}}

First we define some directory paths. 
You should adjust for your particular machine. 
Note, for LHAPDF6, the individual ``dat'' and ``info'' files 
are stored in subdirectories. 

\indent
\pmb{\text{pacDir}=\text{{``}../ManeParse/Demo/packs{''}}}\\
\indent
\pmb{\text{pdfDir}=\text{{``}../LHAPDF{''}}}\\
\indent
\pmb{\text{subDir1}=\text{pdfDir}\lg \text{{``}/MSTW2008nnlo68cl{''}}}\\
\indent
\pmb{\text{subDir2}=\text{pdfDir}\lg \text{{``}NNPDF30\ts nnlo{\ts}as{\ts}0118{\ts}nf{\ts}6{''}}}\\
\indent
\pmb{\text{ctqDir}=}\pmb{\text{{``}../ManeParse/Demo/PDF{\ts}Sets/PDS{''}}} \\

\noindent
Next, we load the \Mane\ packages. 
The \cmd{pdfCalc} package is automatically 
loaded by both  \cmd{pdfParseLHA} and \cmd{pdfParseCTEQ}, so we
do not need to do this separately.

\indent
\pmb{\text{Get}[\text{pacDir}\lg \text{{``}/pdfParseLHA.m{''}}];}\\
\indent
\pmb{\text{Get}[\text{pacDir}\lg \text{{``}/pdfParseCTEQ.m{''}}];}\\
\indent
\pmb{\text{Get}[\text{pacDir}\lg \text{{``}/pdfErrors.m{''}}];}\\

\noindent
\cmd{pdfParseLHA}  will read the PDF set and 
assign  an ``iSet'' number, 
which in this case is $1$.

\indent
\pmb{\text{iSetMSTW}=}\\
\indent
\pmb{\text{pdfParseLHA}[}\\
\indent\qquad \pmb{\text{subDir1}\lg \text{{``}/MSTW2008nnlo68cl.info{''}},}\\
\indent\qquad \pmb{\text{subDir1}\lg \text{{``}/MSTW2008nnlo68cl{\ts}0000.dat{''}}]}\\
\indent 
Out[...]:=\(1\)\\

\noindent
The ``iSet'' numbers are assigned sequentially, and are returned by  
\cmd{pdfParseLHA} which we use to define the variable \cmd{iSetMSTW} (=1 in this example). 
We can then evaluate the PDF values. 

\indent
\pmb{\text{iParton}=0; \text{(*} \text{Gluon} \text{*)}}\\
\indent\pmb{x=0.1;}\\
\indent\pmb{q=10.;}\\
\indent\pmb{\text{pdfFunction}[\text{iSetMSTW},\text{iParton},x,q]} \\
\indent 
Out[...]:=\(11.714\) \\

\noindent
Next, we can read in an NNPDF PDF set.

\indent
\pmb{\text{iSetNNPDF}=}\\
\indent
\pmb{\text{pdfParseLHA}[}\\
\indent \quad 
\scalebox{0.9}[1.0]{\small
\pmb{\text{subDir2}\lg  \text{{``}/NNPDF30{\ts}nnlo{\ts}as{\ts}0118{\ts}nf{\ts}6.info{''}}},
}\\
\indent  \quad
\scalebox{0.9}[1.0]{\small
\pmb{\text{subDir2} \lg   \text{{``}/NNPDF30{\ts}nnlo{\ts}as{\ts}0118{\ts}nf{\ts}6{\ts}0000.dat{''}}}]
} \\
\indent 
Out[...]:=\(2\)\\

\noindent
We can then evaluate this PDF. 
We find it is similar (but not identical) to the value above. 

\indent
\pmb{\text{pdfFunction}[\text{iSetNNPDF},\text{iParton},x,q]}\\
\indent 
Out[...]:=\(11.8288\) \\

\noindent
Finally, we load a \cmd{ctq66} PDF file in the older ``pds'' 
format using the \cmd{pdfParseCTEQ} function; note
this only  takes a single file as the ``info'' details
are contained in the ``pds'' file header.

\indent
\pmb{\text{iSetC66}=\text{pdfParseCTEQ}[}\\ 
\indent \qquad \qquad
\pmb{\text{ctqDir}\lg \text{{``}/ctq66.00.pds{''}}];}\\
\indent
Out[...]:= 3 \\

\indent
\pmb{\text{pdfFunction}[\text{iSetC66},\text{iParton},x,q]}\\
\indent
Out[...]:= 11.0883 \\

\noindent
Now that we have these functions defined inside of Mathematica,
we can make use of all the numerical and graphical functions. 
Detailed working examples are provided in the auxiliary files.

\subsection{\texorpdfstring{$N_F$}{NF}-Dependent PDF Example}
\addtocontents{toc}{\protect\setcounter{tocdepth}{3}}
\addcontentsline{toc}{subsection}{\texorpdfstring{$N_F$}{NF}-Dependent PDF Example}
\addtocontents{toc}{\protect\setcounter{tocdepth}{-1}}

\begin{figure}[t]  
\centering{}
\includegraphics[width=0.45\textwidth]{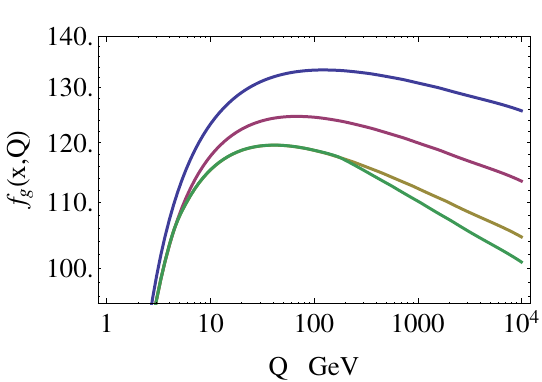}
\protect
\caption{
We display the gluon PDF $f_g (x,Q)$
at $x=0.03$ vs. $Q$ for $N_F=\{3,4,5,6\}$;
$N_F=3$ is the largest, and $N_F=6$ is the smallest curve.
\label{fig:nfPDF}}
\end{figure}

We provide an example of implementing the $N_F$-dependent PDFs 
within the \Mane\ framework using the 
matched set of PDFs\footnote{These PDF sets are available at \cmd{http://ncteq.hepforge.org/}.}  with $N_F=\{3,4,5,6\}$ from Ref.~\cite{Kusina:2013slm}. 
We load the \Mane\ packages as above, and then read in the grid files which are in ``pds'' format. \\

\indent
\pmb{\text{pdfDir}=\text{{``}../vfnsnf{''}};}\\
\indent
\pmb{\text{iSetNF3}=\text{pdfParseCTEQ}[\small{\text{pdfDir}\lg \text{{``}/nf3{\ts}q1.2.pds{''}}}]}\\
\indent
\pmb{\text{iSetNF4}=\text{pdfParseCTEQ}[\small{\text{pdfDir}\lg \text{{``}/nf4{\ts}q1.2.pds{''}}}]}\\
\indent
\pmb{\text{iSetNF5}=\text{pdfParseCTEQ}[\small{\text{pdfDir}\lg \text{{``}/nf5{\ts}q1.2.pds{''}}}]}\\
\indent
\pmb{\text{iSetNF6}=\text{pdfParseCTEQ}[\small{\text{pdfDir}\lg \text{{``}/nf6{\ts}q1.2.pds{''}}}]}\\

\noindent 
\cmd{pdfParseCTEQ} returns the ``iSet'' number and we store these in \{\cmd{iSetNF3, ...}\}.
The below function \cmd{pdfNF}  allows the user to choose  $N_F$, and then  returns 
the appropriate PDF.  \\

\indent
\pmb{\text{Clear}[\text{pdfNF},\text{nf},\text{iParton},x,q];}\\
\indent\pmb{\text{pdfNF}[\text{nf{\ts}},\text{iParton{\ts}},\text{x{\ts}},\text{q{\ts}}]\text{:=}\text{Module}[\{\text{iSet}=0\},}\\\indent \qquad 
\pmb{\text{If}[\text{nf}==3,\text{iSet}=\text{iSetNF3}];}\\ \indent \qquad 
\pmb{\text{If}[\text{nf}==4,\text{iSet}=\text{iSetNF4}];}\\ \indent \qquad 
\pmb{\text{If}[\text{nf}==5,\text{iSet}=\text{iSetNF5}];}\\ \indent \qquad 
\pmb{\text{If}[\text{nf}==6,\text{iSet}=\text{iSetNF6}];}\\ \indent \qquad 
\pmb{\text{If}[\text{iSet}\text{==}0,\text{Return}[\text{Null}]];}\\ \indent \qquad 
\pmb{\text{Return}[\text{pdfFunction}[\text{iSet},\text{iParton},x,q]]}\\ \indent \quad
\pmb{]}\\

\noindent 
Note in the \cmd{pdfNF} function, the ``iSet'' variable is local to the \cmd{Module}. 
We now compute  some sample values. \\

\indent
\pmb{\text{iParton}=0; \text{(*} \text{Gluon} \text{*)}}\\ \indent
\pmb{x=0.03;}\\ \indent
\pmb{q=10.;}\\ \indent
\pmb{\{\text{pdfNF}[3,\text{iParton},x,q],}
\pmb{\text{pdfNF}[4,\text{iParton},x,q],}\\ \indent  \ 
\pmb{\text{pdfNF}[5,\text{iParton},x,q],}
\pmb{\text{pdfNF}[6,\text{iParton},x,q]\}} \\
\indent
Out[...]:=\{123.288, 117.694, 115.331, 115.341\} \\

\noindent
As we have taken $Q=10~$GeV, we are above the charm and bottom transition,
but below the top transition; hence the $N_F=\{5,6\}$ results are the 
same, but the  $N_F=\{3,4\}$ values differ.

In Fig.~\ref{fig:nfPDF} we display the gluon PDF vs. $Q$ for $N_F=\{3,4,5,6\}$.
We observe as we activate more flavors in the PDF evolution
the gluon is reduced as a function of $N_F$. 
This decrease in the gluon PDF will be (partially) compensated by the 
new $N_F$ channels. 


\newpage

\bibliographystyle{elsarticle-num}
\bibliography{mp.bib}







\end{document}